\documentclass[twocolumn,final,natbib]{svjour3}

\usepackage{graphicx}
\DeclareGraphicsExtensions{.ps }

\newcommand {\beq}{\begin{equation}}
\newcommand {\eeq}{\end{equation}} 
\newcommand {\beqr}{\begin{eqnarray}}
\newcommand {\eeqr}{\end{eqnarray}} 
\newcommand{\rf}[1]{(\ref{#1})}

\newcommand{\der}[2]{{\fr{ \textstyle \partial #1}{ \textstyle \partial #2}}}

\newcommand{\fr}[2]{{\frac{ \textstyle #1}{ \textstyle #2}}}

\newcommand {\tm}{\times}
\newcommand{\rfg}[1]{fig.\ref{#1}}

\newcommand{\costfun}{{\cal  I }}

\usepackage{pdftricks}
\usepackage{amsmath}
\usepackage{wasysym}
\usepackage[english]{babel}

\newcommand {\Dx}{\mathbf{D_x}}
\newcommand {\Dy}{\mathbf{D_y}}

\renewcommand {\deg}{^\circ}

\begin{document}
 
\title{Optimal boundary conditions at the staircase-shaped coastlines. }
\author{ Eugene Kazantsev}
\journalname{Ocean Dynamics} 
\institute{ INRIA, projet MOISE, 
 Laboratoire Jean Kuntzmann,
BP 53,
38041 Grenoble Cedex 9, 
France. \email{ kazan@imag.fr} }

%\picstoend

\maketitle

\begin{abstract}
A 4D-Var data assimilation technique is applied to  the rectangular-box  configuration of the NEMO in order to identify the optimal parametrization of  boun\-dary conditions at  lateral boundaries. The case of the staircase-shaped coastlines is studied by rotating the model grid around the center of the box.   It is shown that, in some cases, the formulation of the boundary conditions at the exact boundary leads to appearance of   exponentially growing modes while optimal boundary conditions allow to correct the errors induced by  the staircase-like appriximation of the coastline.   

\end{abstract}

%\begin{keyword} 
{\bf Keywords: }{\it Variational Data Assimilation; Boundary conditions; NEMO model.}
%\end{keyword}
 
%Started the 27.11 2013

%Submitted  the 

\section{Introduction}

Contemporary ocean general circulation  models are usually discretized by finite differences on a grid with rectangular cells. This discretization simplifies  the model and accelerates its integration in the middle of the ocean. In the same time, it can not provide an accurate approximation of the  model boundary. Ocean coastlines are usually approximated by some staircase-like structures that neither respect the correct boundary position, nor the normal and tangential directions. This problem ari\-ses as at the bottom of the ocean and at the lateral boundary as well. 

This paper is devoted to the analysis of the lateral boundary conditions in frames of the rectangular box configuration of the Nemo model (see \cite{madec-nemo}) in the case when the boundary is not aligned with the model grid.  The interest of this study is determined by the difficulties of correct representation of the Gulf Stream separation from the continent. In the extended review of the present knowledge on this question (\cite{chassignet}), it is noted that the separation
mechanism in numerical models remains very sensitive to choices made in the numerical model for parameterizations of the dissipation and of the boundary layer. Numerous papers are devoted to the analysis and comparison of  different numerical schemes that approximate physical mechanisms near the step-like coastlines. One can cite the paper of \cite{Adcroft98} in which the solution of a shallow-water model is analyzed in a square box for free-slip and no-slip conditions. It is shown in this paper, that  inappropriate approximation of the lateral friction term may lead to underestimated or overestimated boundary stress, resulting sometimes in the replacement of  free-slip boundary conditions by no-slip ones.

In \cite{dupont03}, the influence of approximations of the lateral dissipation is studied together with the approximation of the advection term showing that some conventional discretizations of this term are also inappropriate and may lead to instabilities. 

On the example of inviscid linear shallow-water mo\-del,  it is shown in \cite{griffiths13} that  staircase-like boundary reduces the approximation of the  Kelvin wave speed down to the first order in $h$ and degrades the accuracy of numerical simulation of physical phenomena. 

However, even if we follow the recommendations of \cite{Adcroft98, dupont03} and use enstrophy conserving advection scheme together with the "vorticity-divergence" form of the viscous stress tensor, the influence of the   staircase-like boundary is still important in a full physics model. In this paper we shall analyze  western boundary currents produced by the Nemo model in the presence and in the absence of the staircase.   The same configuration as in  papers cited above is used: the model is integrated in a rectangular box, but the model grid can be rotated on a desired angle around the center of the rectangle. This rotation ensures different alignment of the physical boundary and the  grid while preserving all other dynamical properties of the model.

Several  techniques are available  to implement the complex conditions at immersed boundaries. One of them consists in using a finite element method that can treat the complex boundary in a more accurate way (see, e.g.     
 \cite{DanilovFEM, yakovlevFEM}). However, up to now, finite element methods are only used to discretize ocean models in the most difficult geometries  and finite differences models dominate due to their  relative simplicity and robustness.

Another way to approximate the boundary is to use a shaved-cell technique (see, e.g., \cite{Adcroft97, partsteps} ), which can be viewed as a method
where the variables are brought to the boundaries by means of extrapolation/interpolation. However, this me\-thod has several disadvantages also: interpolations may have to be adapted to ensure
stability, accurate discretization of the model equations on cut cells is not trivial, and special measures
may be required to deal with very small cut cells. Moreover, as it is shown in this paper, application of piecewise shaved cells to approximate lateral boundaries may result in unstable advection operator. 

Alternative approach, discussed  in this paper, suppose to ask the model's opinion about  "optimal" lateral boundary conditions  that will be able to compensate the error committed by the staircase-like approximation of the coastline by the model grid. The word "optimal" is used here in the sense of the 4D-Var data assimilation: what  boundary approximation would realize the minimum of the specially constructed cost function. 

We follow the procedure proposed in \cite{orca-paper} and, instead of controlling the shape of the coast or  boundary conditions themself, we  control  the  numerical scheme  that is used to approximate differential operators in points adjacent to  the lateral boundary. 
As it has been noted in  \cite{Leredde}, particular attention must be paid to the discretization technique  which must respect several rules because it is the discretization of the model's operators takes into account the set of boundary conditions and introduces them into the model.  Indeed, boundary conditions participate  in discretized operators, but considering the discretization itself, we take into account also the way they influence the model.  

The purpose of this paper is to analyze the influence of lateral coastline  on the model flow in a simple, academic configuration of a full-physics model and to propose a way of optimization of the approximation of the coastline and boundary conditions by variational data assimilation.  

As well as in \cite{orca-paper}, tangent linear and adjoint   codes, necessary for variational data assimilation,  have been obtained by the AD Tapenade described in \cite{Hascoet04}. Particular utility of automatic differentiation in the case of boundary conditions control is related to the fact that the derivative of the model with respect to boundary conditions is two or three times longer (as well as in terms of the development, the number lines of the code and the necessary CPU time) than the derivative used to control the initial point of the model (see \cite{sw-lin} for details).

A rectangular box configuration of the Nemo \footnote{http://www.nemo-ocean.eu/} (the Nucleus for European Modelling of the Ocean) is discussed here in frames of twin experiments. Pseudo-obser\-vational  data are generated by the model on the grid aligned with boundaries. These data are assimilated into the same model, but on the grid rotated on some angle. 

Two classical wind forcings are considered in the paper: single and double gyre wind stress at the surface of the ocean. The first one provides an almost stationary solution with strong currents near the Western and the Northern boundaries allowing us to analyze the influence of the staircase shaped approximation of the coastline on the flow in the boundary layer. The second one leads to a formation of the jet stream in the middle of the rectangle and allows us to simulate the separation of the Western boundary currents.

\section{Rectangular-box configuration on the Nemo}

\begin{figure*}[t]
\hrule
\beqr
\der{u}{t}&=&
 \underbrace{
 \overline{\overline{v}^x}^y \overline{(\omega+f)}^y- 
 \delta_x\left[\fr{ \overline{u^2}^x+  \overline{v^2}^y}{2}\right]
 -\overline{\biggl( \overline{w}^x \delta_z\left[ u\right]\biggr)}^z
 }_{\mbox{ Advection}}+
 \underbrace{
   \delta_x\left[{A^{xy}_u\xi}\right]-{ \Dy}{A^{xy}_u\omega}
 }_{\mbox{ Lateral diffusion}}   +
 \underbrace{
   g\int_{0}^z\delta_x\left[  \overline{\rho(x,y,\zeta)}^z \right]d\zeta
 }_{\mbox{Hydrostatic  Press. grad.}}  +   
 \underbrace{  
   \delta_z\left[ A^z_u \delta_z\left[u\right]\right]
 }_{\mbox{ Vertical diffusion}}   + 
  \nonumber\\ &+& 
 \underbrace{ 
 g \delta_x\left[(\eta+T_c\partial_t\eta)\right]
  }_{\mbox{ Surface Pres. Grad.}}   
 \label{1.1}\\ 
 \der{v}{t}&=&
 \underbrace{
  \biggl(- \overline{\overline{ u}^x}^y\biggr)\overline{(\omega+f)}^x- \delta_y\left[\fr{\overline{ u^2}^x+ \overline{ v^2}^y}{2}\right]-\overline{\biggl(\overline{ w}^x \delta_z\left[ v\right]\biggr)}^z
 }_{\mbox{ Advection}}+
 \underbrace{
   \delta_y\left[{A^{xy}_v\xi}\right]+{ \Dx}{A^{xy}_v\omega}
 }_{\mbox{ Lateral diffusion}}   +
 \underbrace{
   g\int_{0}^z\delta_y\left[ \overline{ \rho(x,y,\zeta)}^z\right] d\zeta
 }_{\mbox{Hydrostatic  Pressure grad.}}  + 
 \nonumber\\ &+&   
 \underbrace{  
   \delta_z\left[ A^z_v \delta_z\left[v\right]\right]
 }_{\mbox{ Vertical diffusion}}   + 
 \underbrace{ 
 g \delta_y\left[(\eta+T_c\partial_t\eta)\right]
  }_{\mbox{ Surface Pres. Grad.}}   
 \label{1.2}\\ 
 \der{T}{t}&=&
  \underbrace{
   -\delta_x\left[ (u\overline{ T}^x)\right]-\delta_y\left[ (v \overline{ T}^y)\right]-\delta_z\left[ (w \overline{ T}^z)\right]
     }_{\mbox{Advection}}   +
 \underbrace{ 
   A^{xy}_T\biggl(\delta_x\left[\delta_x\left[T\right]\right]+\delta_y\left[\delta_y\left[T\right]\right]\biggr)
  }_{\mbox{Lateral diffusion}}     +
  \underbrace{
    \delta_z\left[ A^z_T\delta_z\left[ T\right]\right]
  }_{\mbox{Vert. diffusion}}       
    \label{1.3} \\
 \der{s}{t}&=&
  \underbrace{
   -\delta_x\left[ (u\overline{ s}^x)\right]-\delta_y\left[ (v \overline{ s}^y)\right]-\delta_z\left[ (w \overline{ s}^z)\right]
     }_{\mbox{Advection}}   +
 \underbrace{ 
   A^{xy}_T\biggl(\delta_x\left[\delta_x\left[s\right]\right]+\delta_y\left[\delta_y\left[ s\right]\right]\biggr)
  }_{\mbox{Lateral diffusion}}     +
  \underbrace{
    \delta_z\left[ A^z_T\delta_z\left[ s\right]\right]
  }_{\mbox{ Vertical diffusion}}       
   \label{1.4} \\
  \der{\eta}{t}&=&   w(x,y,z=\mbox{surface}) \label{1.5} \\
 \xi&=& \delta_x\left[u\right]+\delta_y\left[v\right],\hspace{3mm}
 \omega={\Dy}u-{\Dx}v,
 \label{1.6}\\ 
 w&=&\int_{bottom}^z \xi(x,y,\zeta) d\zeta,\hspace{3mm}
 \rho=\rho(T,s)\label{1.7}
 \eeqr
 \end{figure*}  
  
Rectangular box of $30\deg$ in longitude and $20\deg$ in latitude is considered with $0.25\deg$ resolution in both directions.  In the vertical direction the total depth of $4195$m is separated into 4 layer of equal thickness. The model grid is composed of $120\tm 80\tm 4$ nodes for each variable. This configuration is similar to the conventional SEABASS configuration of the Nemo. 
So far, the attention is focused on the influence of lateral boundaries,  the bottom is supposed to be flat.  

Spatially discretized equations of the model are written in the system \rf{1.1}--\rf{1.7}, in which  
 operators $\delta\left[ \cdot \right]$ and $\overline{\cdot}$ represent conventional second order approximations of  derivatives and interpolations on the Arakawa C-grid.   Interpolations are calculated as a  weighted mean of two function values at the adjacent nodes. Weights are defined to be proportional to the grid steps of corresponding cells in order to achieve the second order interpolation of a grid function (see  \cite{madec-nemo}). However, we write  these operators in a simplified way, assuming both the argument and the result of the interpolation  and the derivative are  multiplied by an appropriate weight:
 \beq
 \overline{w}^z=\fr{w_{i,j,k+1}+w_{i,j,k}}{2}, \quad
 \delta_y\left[ u \right]=\fr{u_{i,j+1,k}-u_{i,j,k}}{h} \label{s-mid}
\eeq

 Operators $\Dx, \Dy$ (in equations \rf{1.1},\rf{1.2},\rf{1.6}) approximate derivatives also: they coincide with $\delta\left[ \cdot \right]$ in the middle of the ocean, but they may differ from conventional derivatives near the boundary. 
 
 The set of variables in this system  consists of the following: $u,v$ and $w$ ---  zonal, meridional and vertical velocity components, $T$ and $s$ --- the potential temperature and salinity;   $\xi$ and $\omega$ --- horizontal divergence and vorticity; $\eta$ -- the sea surface elevation; $\rho$ --- the density anomaly that is defined as a function of the temperature and salinity by  the state equation.  As one can see, $u,v,T,s,\eta$ are prognostic variables while  $w, \xi, \omega$ and $\rho$ are diagnostic ones.

 Among other parameters in these equations, one can see  the Coriolis parameter  $f=2\Omega\sin(\phi)$. the  gravity acceleration $g=9.81\fr{m}{s^2}$, lateral diffusion coefficients  $A^{xy}_u=A^{xy}_T=200\fr{m^2}{s}$,  coefficients of the vertical diffusion $A^z_u=A^z_v=1.2\tm 10^{-4}\fr{m^2}{s},\;\; A^z_T=A^z_s=1.2\tm 10^{-5}\fr{m^2}{s}$. 
 
The term  $T_c\partial_t\eta$ in  equations \rf{1.1} and \rf{1.2} is introduced to dump the external gravity waves. These waves are fast so their timescale is short with respect to other processes described by the primitive equations. Explicit resolution of these waves requires an excessively small time step, which is not needed to resolve other physical processes supported by the equations. 
 Consequently,   the filter of temporally unresolved external gravity waves, proposed in \cite{RoulletMadec}, is introduced into the model. The cutoff time  $T_c$ is equal to  one time step of the model.  

The purpose of this paper consists in the variational analysis of  the influence of the staircase-like boundary on the model solution. for this, we formulate several model configurations, all of them are based on a  rectangular box, aligned with the latitude-longitude coordinates and placed  in the mid-latitude region.   The model grid  may be either aligned with the coordinate system (and, hence, with the  walls of the box), or inclined  at some angle  to coordinates. This configuration has already been used in numerous papers for similar purposes (see, for example \cite{Adcroft98,dupont03,griffiths13}), but now we can benefit more of this set of configurations, assimilating the information about the solution obtained on the aligned grid into the model on inclined grid with staircase-like boundaries. 

The box is  placed in the mid-latitude  region, between $\phi=24\deg$ and $\phi=44\deg$.  The model is forced by the stationary  wind stress with now classical single or double gyre pattern. 
\beq
\tau_\phi=0, \quad \tau_\lambda=-0.1\fr{N}{m^2}\tm \fr{\cos(n\pi(\phi-24\deg))}{44\deg-24\deg} \label{forc}
\eeq
where $n$ is either $1$ in the single gyre experiments, or $2$ in the double gyre ones. The wind stress influences the model solution by the boundary conditions on the surface of the ocean, prescribed for the vertical diffusion operator $\delta_z\left[ A^z \delta_z\left[ \cdot \right]\right]$:

\beq
\der{u}{z}\bigg\vert_{surface}=\fr{\tau_x}{hz_1\rho_0},\quad
\der{v}{z}\bigg\vert_{surface}=\fr{\tau_y}{hz_1\rho_0},\nonumber 
\eeq
where $\tau_x$ and $\tau_y$ are components of the wind stress on the model grid that may be inclined at angle $\varphi$ to the coordinate system:
\beq
\tau_x=\tau_\lambda \cos(\varphi)-\tau_\phi\sin(\varphi), \quad 
\tau_y=\tau_\lambda \sin(\varphi)+\tau_\phi\cos(\varphi)
\eeq

The model is discretized on the grid, which represents the generalization to three dimensions of the well-known ``C'' grid in 
Arakawa's classification \cite{Mesinger_Arakawa_Bk76}. The arrangement of variables is the same in all directions. 
It consists of cells centered on scalar points ($T$, $s$, $\eta$, $\rho$) with vector 
points $(u, v, w)$ defined at the center of each face of the cells. The relative and 
planetary vorticity, $\omega$ and $f$, are defined at the center of each vertical edge.

Following \cite{assimbc1}, \cite{sw-nl}, instead of controlling physical boundary conditions, we use    more general framework controlling the way boundary conditions are introduced in the model operators. Thus,  expressions for derivatives $\Dx, \Dy$, are modified at the grid-nodes adjacent to the boundary, i.e. near the continents. This allows us to control simultaneously the boundary conditions, the position of the boundary with respect to the grid and the possible interpolation of the boundary conditions to the grid. 

Let us suppose the index $i=0$ corresponds to the left rigid  boundary as it is shown in \rfg{xgrid}. That implies the index $i=1/2$ corresponds to the first $v$-point in the ocean and the index $i=1$ corresponds to the first $\omega$-point in the ocean. To approximate $\der{v}{x}$ and $\der{\omega}{x}$ at these points we must take into account the boundary conditions prescribed for $v$ and for $\omega$ and write either  
$$\der{v}{x}\bigg\vert_{0}=0,\quad 
\der{\omega}{x}\bigg\vert_{1/2}=\fr{\omega_1}{h}$$
 in the case of  free-slip condition $\omega_0=0$, 
or  
$$\der{v}{x}\bigg\vert_{0}=\fr{2v_{1/2}}{h}, \quad 
\der{\omega}{x}\bigg\vert_{1/2}=\fr{\omega_1-\omega_0}{h}$$
  in the case of  no-slip condition $v_0=0$.

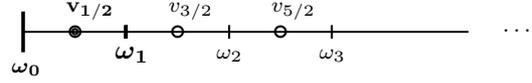
\begin{figure}[h]
\setlength{\unitlength}{0.9mm}
\newcount\indi
\newcount\num
\begin{picture}(80,15)
\linethickness{0.25mm}
\put(5,10){\line(1,0){65}}
\put(75,10){\ldots}
\put(5,7){\line(0,1){6}}\put(4.75,7){\line(0,1){6}}
\thicklines
%%%%%%%%%%%%%%%%%%%%%%%%%%%%%%%%%%%%%%%
%U velocity (integer points)
%%%%%%%%%%%%%%%%%%%%%%%%%%%%%%%%%%%%%%%

\put(20,9){\line(0,1){2}}\put(19.75,9){\line(0,1){2}}
\multiput(35,9)(15,0){2}{\line(0,1){2}}
\put(2,4){  $\mathbf{\boldsymbol\omega_{0}}$ }
\indi=1
\multiput(32,6)(15,0){2}{ 
\global\advance\indi by 1 $\omega_{\the\indi}$ }

%%%%%%%%%%%%%%%%%%%%%%%%%%%%%%%%%%%%%%%
%h height
%%%%%%%%%%%%%%%%%%%%%%%%%%%%%%%%%%%%%%%
%\multiput(12,9)(15,0){3}{\line(1,1){2}}
%\multiput(14,9)(15,0){3}{\line(-1,1){2}}
\multiput(12.5,10)(15,0){3}{\circle{1.5}}
\indi=0

\multiput(25,13)(15,0){2}{ 
\global\advance\indi by 1 \num=\indi \multiply\num by 2 \global\advance\num by 1 $v_{\the\num/2}$ }

\put(12.5,10){\circle{0.75}}
\put(10,13){  $\mathbf{v_{1/2}}$ }
\normalsize
\put(20,9){\line(0,1){2}}
\put(17,6){  $\mathbf{\boldsymbol\omega_{1}}$ }
\end{picture} 
\caption{ Structure of the horizontal grid. }
%\refstepcounter{fig}
\label{xgrid}
\end{figure}

  To control these conditions together with their approximations in the model, we  approximate them using a more general formula
\beqr
(\Dx v)_{0,j,k}&=& \fr{ \alpha^{Dxv^l}_0 v_{1/2,j,k} }{h}\nonumber\\
(\Dx v)_{1,j,k}&=& \fr{ \alpha^{Dxv^l}_1 v_{1/2,j,k}+\alpha^{Dxv^l}_2 v_{3/2,j,k} }{h}\nonumber\\
(\Dx \omega)_{1/2,j,k}&=&\fr{ \alpha^{Dx\omega^l}_1 \omega_{0,j,k}+\alpha^{Dx\omega^l}_2 \omega_{1,j,k} }{h}
\label{s-1}
\eeqr
where appropriate values are given to coefficients $\alpha$ corresponding to prescribed boundary conditions ( for example, 
\beqr
&& \alpha^{Dxv^l}_0=0,\; 
\alpha^{Dxv^l}_1=-1,\; 
\alpha^{Dxv^l}_2 =1, \nonumber\\
&&\alpha^{Dx\omega^l}_1=0,\; 
\alpha^{Dx\omega^l}_2=1
\label{alpha-initguess}
\eeqr when  free-slip conditions are prescribed). 

Value ofr $(\Dx v)_{N,j,k}, (\Dx \omega)_{N-1/2,j,k}, \ldots$ near the right boundary are calculated by  similar expressions, but with different coefficients $\alpha^{Dxv^r}_0, \alpha^{Dxv^r}_1, \ldots$:
\beqr
(\Dx v)_{N,j,k}&=& \fr{ \alpha^{Dxv^r}_0 v_{N-1/2,j,k} }{h}\nonumber\\
(\Dx v)_{N-1,j,k}&=& \fr{ \alpha^{Dxv^r}_1 v_{N-1/2,j,k}+\alpha^{Dxv^r}_2 v_{N-3/2,j,k} }{h}\nonumber\\
(\Dx \omega)_{N-1/2,j,k}&=&\fr{ \alpha^{Dx\omega^r}_1 \omega_{N,j,k}+\alpha^{Dx\omega^r}_2 \omega_{N-1,j,k} }{h}
\label{rightbnd}
\eeqr
The same technique is used to approximate operators $\der{u}{y}\sim \Dy u$ and $\der{\omega}{y}\sim \Dy \omega$ near the Northern and near the Southern boundaries.

Coefficients $\alpha$ play the role of  control variables in this paper. Operators  $\Dx, \Dy$ are allowed to change their properties near the boundaries in order to find the best fit with requirements of the model and data.   To specify optimal values of  control variables   we  perform the data assimilation procedure.

Taking into account that at different points the optimal boundary conditions may differ, we
have to consider the spatially variable coefficients.  Thus, $\alpha^{Dx}$ and  $\alpha^{Dy}$ are  allowed to vary from one boundary point to another. 

In the paper \cite{orca-paper}, boundary conditions for all derivatives and interpolation operators have been controlled. There has been shown that only few of them  influence the model solution significantly. In this paper, we have tried to control all the horizontal operators and have selected the boundary conditions for $u,v$ and $\omega$ only, because they have the largest impact on the solution in this configuration. More precisely, boundary conditions are controlled only in the calculation of the relative vorticity and in the vorticity part of the horizontal dissipation of $u$ and $v$.

 In order to distinguish the controlled derivatives from all others approximation of derivatives $\delta\left[ \cdot \right]$ ,   we denote them as  $\Dx, \Dy$ in  equations \rf{1.1}, \rf{1.2}, \rf{1.6}. It should be noted that everywhere $\Dx\cdot =\delta_x\left[ \cdot \right]$, $\Dy\cdot =\delta_y\left[ \cdot \right]$, except near the boundary.

Total set  of control coefficients $\alpha$ counts about  20 000 elements while the dimension of the model state is more than 160 000.

\section{Data assimilation.}

Academic configuration of the model  allows us to  work in frames of  twin experiments. The model was spun up during 1000 days on the grid aligned with the boundary (grid lines are either parallel or perpendicular to the coast) starting from the  state:
\beqr
u=0,\;v=0,\;\eta=0,\;s=35.5\permil,\nonumber\\
T=25^\circ+24.1^\circ\tm (e^{-z/800}-1)\nonumber
\eeqr

 The final state of the spin-up is used as the starting point to produce the artificial observational data set. The model on the aligned grid is started from this state and forwarded in time  for 100 days. During this run, variables $u, v$ and $\eta$   have been interpolated to the rotated  grid at each time step and the result is saved to be used in the assimilation experiments as $(u^{obs}, v^{obs}, \eta^{obs})$.

The same  final state of the spin-up has also been  interpolated to the rotated grid and used both as the background state and as the first guess of the initial conditions for the model on the rotated grid. Looking for optimal $\alpha$, we use  classical combinations (like \rf{alpha-initguess}) both as the first guess and as the background.

We perform bicubic interpolation to get  the variables values at nodes of the rotated grid in order to get smoother and more precise  fields. However, even bicubic algorithm  occurs to be too inaccurate for the interpolation of the velocity fields. In fact, $\der{u}{x}$ and $\der{v}{y}$ almost compensate each other in the divergence operator \rf{1.6} and even small uncorrelated errors in $u$ and $v$  strongly disbalance the divergence and lead to unphysical values of the vertical velocity $w$ \rf{1.7}. In order to keep the balance of the divergence components, we calculate first the relative vorticity and the divergence on the aligned grid, interpolate these values to the rotated grid and reconstruct the velocity fields on the rotated grid from  interpolated vorticity and divergence. In the reconstruction we have to solve a linear system with $N$ variables and $N+1$ equations which has an unique solution under condition $\sum_{i,j} \biggl(\der{u}{x}+\der{v}{y}\biggr)_{i,j} =0 $. But, thanks to impermeability condition imposed on the whole boundary, this condition is always satisfied. Obtained velocity fields ensure rather well compensated divergence and reduce  unphysical effects in the vertical velocity. Only a limited noise remains present in the model solution. 

To completely avoid the influence of  interpolation errors on the model solution,   we perform the joint control of both initial state and boundary parameters $\alpha$ in all experiments below. Taking into account that the adjoint model, constructed for controlling the boundary parameters, contains the whole block necessary to  control  initial conditions, we perform the joint control at almost no additional computational cost.

Thus, in the following experiments we assimilate an interpolated solution of the model on the aligned grid into the model on the rotated grid. Both data and model state are defined at nodes of the same rotated grid, consequently, the observational operator is equal to identity. The only noise as in the data and in the background is due to  interpolation errors, i.e. both data and background contain an uncorrelated noise of the same nature. This fact allows us to use identity covariance matrices simplifying the cost function.  

The cost function we use in all the assimilation experiments below has a form
 \beqr
\costfun&=&\costfun(u\mid_{t=0},v\mid_{t=0},T\mid_{t=0},s\mid_{t=0},\eta\mid_{t=0},\alpha) =\nonumber \\
&=&10^{-4}\biggl(\sum (\alpha-\alpha^{bgr})^2+\int_{xy} \biggl[(u\mid_{t=0}-u^{bgr})^2 +
\nonumber \\
&&+(v\mid_{t=0}-v^{bgr})^2+ (T\mid_{t=0}-T^{bgr})^2 +
\nonumber \\
&&+ (s\mid_{t=0}-s^{bgr})^2+(\eta\mid_{t=0}-\eta^{bgr})^2\biggr] dx\, dy \biggr)+
\nonumber \\
&+& \int\limits_0^T  t \int_{xy} \biggl[(u-u^{obs})^2 +(v-v^{obs})^2+
\nonumber \\
&&+ (\eta-\eta^{obs})^2\biggl] dx\, dy\, dt \label{costfn}
\eeqr
where $(u,v,\eta)$ are the velocity components and SSH in the solution of  the model on the rotated grid at time $t$ obtained starting from $(u\mid_{t=0}, v\mid_{t=0}, T\mid_{t=0}, s\mid_{t=0}, \eta\mid_{t=0})$ with the discretizations of horizontal operators $\Dx, \Dy$  in the vicinity of  continents defined by the set of $\alpha$.

Tracers ($T$ and $s$) are not included in the set of  data to be assimilated. The background term in the cost function is  intentionally weighted by a small coefficient. Having ``observational" data at any time and at each grid-point, the background is important only to control the initial condition for tracers. For all dynamical variables, the  observational part of the cost function dominates.

Taking into account the results obtained in \cite{sw-lin}, we define the cost function  
that gives a larger weight to the difference ``model--observations" at the end of the assimilation window. Since we perform the data assimilation in order to make a forecast, we need a "better" estimate of the model state at the end of the assimilation window because this state is used as the initial point for the forecasting  that starts after the assimilation. For this purpose, we force the model trajectory to go closer to observations at the end of the assimilation window increasing the weight $t$ of the distance in the observational part of the cost function. 

To search for a minimum of the cost function, we  use its gradient with respect to  control parameters which development is described in \cite{orca-paper}. 
The  minimization procedure  described in \cite{lemarechal} is used in all experiments.  This procedure uses the gradient of the cost function in  the limited memory quasi-Newton method.

Tangent and adjoint models have been automatically generated by the Tapenade software presented in \cite{Hascoet04}.  This software analyzes the source code of a nonlinear model and produces the codes of the tangent model  and of its adjoint.  Advantages, shortcoming of the  automatic tangent and adjoint code generation is described in details also in \cite{orca-paper}, as well as the  adjoint code optimization that is necessary to avoid an excessive memory usage.

\section{Single gyre forcing, $45^\circ$ rotation}

We start experiments from the simplest possible case that should clearly distinguish the phenomena related to the staircase-like representation of the boundary and problems induces by this representation.  

The model is forced by the single-gyre wind stress ($n=1$ in \rf{forc}) and subjected to the 
 impermeability and free-slip boundary conditions on the velocity components:
\beq 
\vec{U}\cdot\vec n=0,\quad \der{\vec{U}\cdot\vec{\tau}}{\vec{n}}=0, \label{vecbc}
\eeq
where $\vec{U}=(u,v)$ is the two-dimensional  velocity vector, $\vec n$ and $\vec{\tau}$ are normal and tangential directions with respect to the boundary and $\cdot$ denotes the scalar product.  When the grid is aligned with the boundary, $u$ and $v$ represent either normal or tangential velocity components and the boundary conditions become
\beqr
&&\biggl(u=0,\;\der{v}{x}=0\biggr)_{\mbox{East,West}}, \nonumber \\
&&\biggl(v=0,\;\der{u}{y}=0\biggr)_{\mbox{North,South}}
\eeqr
This implies immediately that the normal  velocity component  and the relative vorticity must vanish everywhere on the boundary. 

If we turn the grid on $45^\circ$ we get a staircase-like approximation of the $45^\circ$-slanted (with respect to the grid) straight line as shown in \rfg{45grid}. One can easily understand that this approximation automatically modifies boundary conditions: impermeability becomes imposed at each $u$ and $v$ nodes on the boundary ($u_{i,i}$ and $v_{i,i}$ in \rfg{45grid}). Indeed, the flux determined by either  $u$ or $v$ velocity components face a perpendicular segment of the stair, obeying the prescribed impermeability condition.   Free-slip condition is taken into account by imposing vanishing relative vorticity at  $\omega_{i,i}$ points   without accepting non-zero tangential flow.

\begin{figure}[h]
\setlength{\unitlength}{0.7mm}
\newcount\indi
\newcount\indj
\newcount\num
\small
\begin{center}
\begin{picture}(80,80)
\linethickness{0.4mm}
%\put(5,5){\line(1,0){20}}
\put(5,25){\line(1,0){20}}
\put(25,25){\line(0,1){20}}
\put(25,45){\line(1,0){20}}
\put(45,45){\line(0,1){20}}
\put(45,65){\line(1,0){20}}
\put(65,65){\line(0,1){10}}

\put(5,5){\line(0,1){20}}
%\put(25,0){\line(0,1){5}}

\multiput(4.7,15)(0.1,0){6}{\line(1,1){64}}
%\multiput(4.7,15)(0.1,0){6}{\line(1,-1){20}}

\linethickness{0.2mm}

\multiput(15,5)(20,0){4}{\line(0,1){80}}
\multiput(0,15)(0,20){4}{\line(1,0){80}}

%\multiput(25,5)(20,0){3}{\circle{1.5}}
\multiput(25,25)(20,0){3}{\circle{1.5}}
\multiput(45,45)(20,0){2}{\circle{1.5}}

\indi=0\indj=0
%\multiput(19,6.5)(20,0){3}{ \global\advance\indi by 1  $\omega_{\scriptscriptstyle\the\indi\the\indj}$ }
\indi=0\indj=1
\multiput(22,22)(20,0){3}{\global\advance\indi by 1  $\omega_{\scriptscriptstyle\the\indi\the\indj}$ }
\indi=1\indj=2
\multiput(42,42)(20,0){2}{ \global\advance\indi by 1 $\omega_{\scriptscriptstyle\the\indi\the\indj}$ }
\indi=2\indj=3
\multiput(62,62)(20,0){1}{ \global\advance\indi by 1 $\omega_{\scriptscriptstyle\the\indi\the\indj}$ }

\multiput(4,14)(20,0){4}{  \line(0,1){2} }
\multiput(24,34)(20,0){3}{  \line(0,1){2} }
\multiput(44,54)(20,0){2}{  \line(0,1){2} }

\indi=-1\indj=0
\multiput(-1,16)(20,0){4}{\global\advance\indi by 1  $u_{\scriptscriptstyle\the\indi\the\indj}$ }
\indi=0\indj=1
\multiput(18,36)(20,0){3}{ \global\advance\indi by 1 $u_{\scriptscriptstyle\the\indi\the\indj}$ }
\indi=1\indj=2
\multiput(39,56)(20,0){2}{\global\advance\indi by 1  $u_{\scriptscriptstyle\the\indi\the\indj}$ }

%\multiput(12,5)(20,0){4}{  \line(1,0){3} }
\multiput(12,25)(20,0){4}{  \line(1,0){3} }
\multiput(32,45)(20,0){3}{  \line(1,0){3} }
\multiput(52,65)(20,0){2}{  \line(1,0){3} }

\indi=-1\indj=0
%\multiput(9,6)(20,0){4}{\global\advance\indi by 1   $v_{\scriptscriptstyle\the\indi\the\indj}$ }
\indi=-1\indj=0
\multiput(9,26)(20,0){4}{\global\advance\indi by 1   $v_{\scriptscriptstyle\the\indi\the\indj}$ }
\indi=0\indj=1
\multiput(29,46)(20,0){3}{\global\advance\indi by 1   $v_{\scriptscriptstyle\the\indi\the\indj}$ }
\indi=1\indj=2
\multiput(49,66)(20,0){2}{\global\advance\indi by 1   $v_{\scriptscriptstyle\the\indi\the\indj}$ }

\indi=-1
\multiput(3,-2)(20,0){4}{ \global\advance\indi by 1 $\the\indi$ }
\indi=-1
\multiput(-2,3)(0,20){4}{ \global\advance\indi by 1 $\the\indi$ }

\end{picture} 
\end{center}
\caption{$45^\circ$ rotated grid}
\label{45grid}
\end{figure}
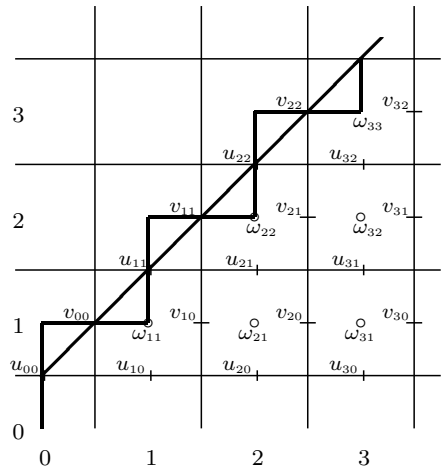

Of course,  altered boundary conditions modify the model solution.   To show the difference between solutions on the aligned grid and on the rotated one, we run the model on both grids for 800 days  waiting for all transition processes are completed and plot two sea surface height (SSH) patterns  on the 800th day. Solutions both on the aligned and rotated grids are almost stationary with a small waves activity. But the boundary current near the North coast is more than two times shorter on the rotated grid (\rfg{sg-ref}B)  than on the aligned one (\rfg{sg-ref}A).

\begin{figure*}[t]
  \begin{center}
  \begin{minipage}[l]{0.48\textwidth}
   A. \\ 
  \centerline{\includegraphics[angle=0,width=0.99\textwidth]{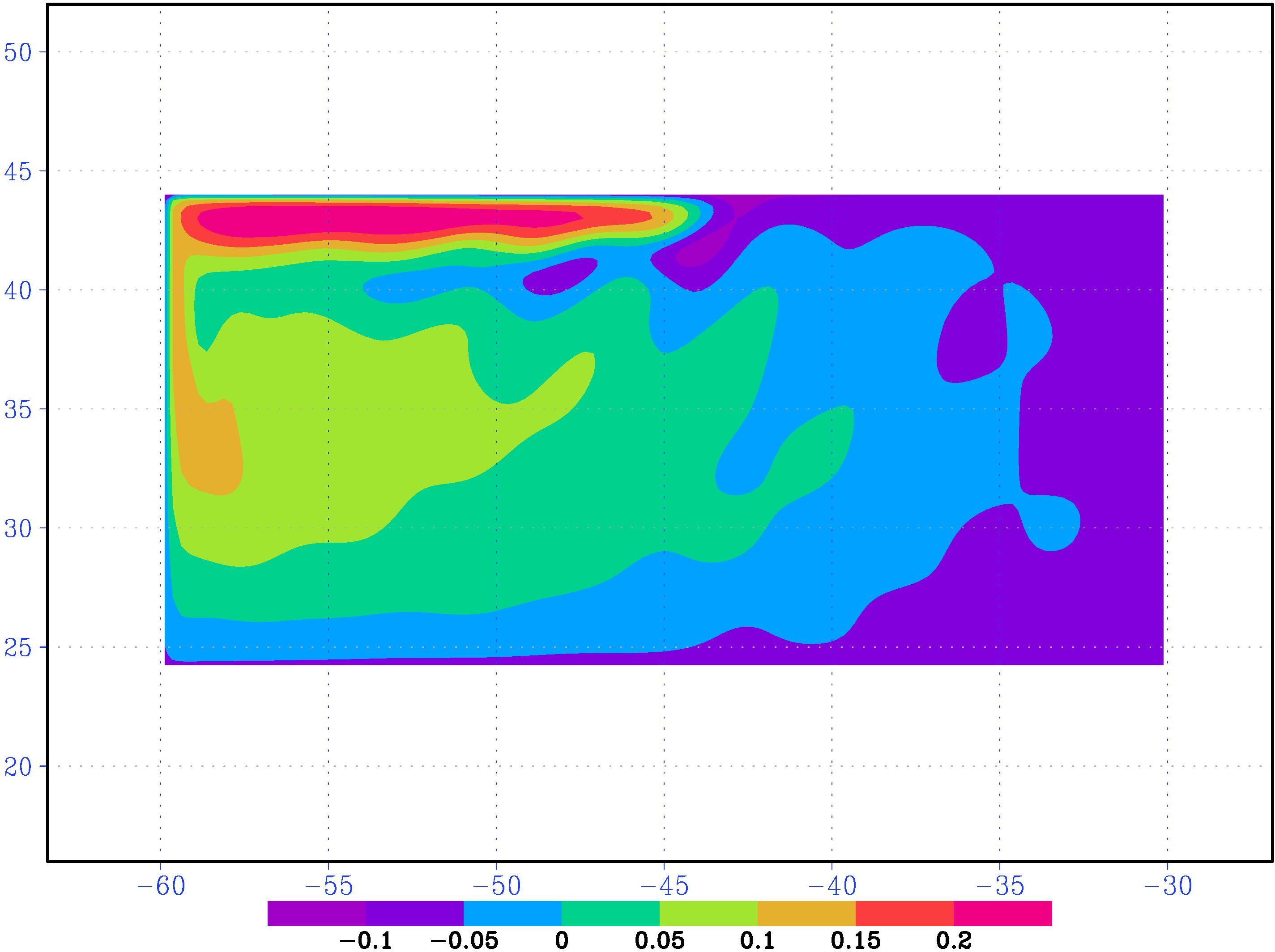}}
  \end{minipage} 
  \begin{minipage}[r]{0.48\textwidth} 
      \hfill B. 
  \centerline{\includegraphics[angle=0,width=0.99\textwidth]{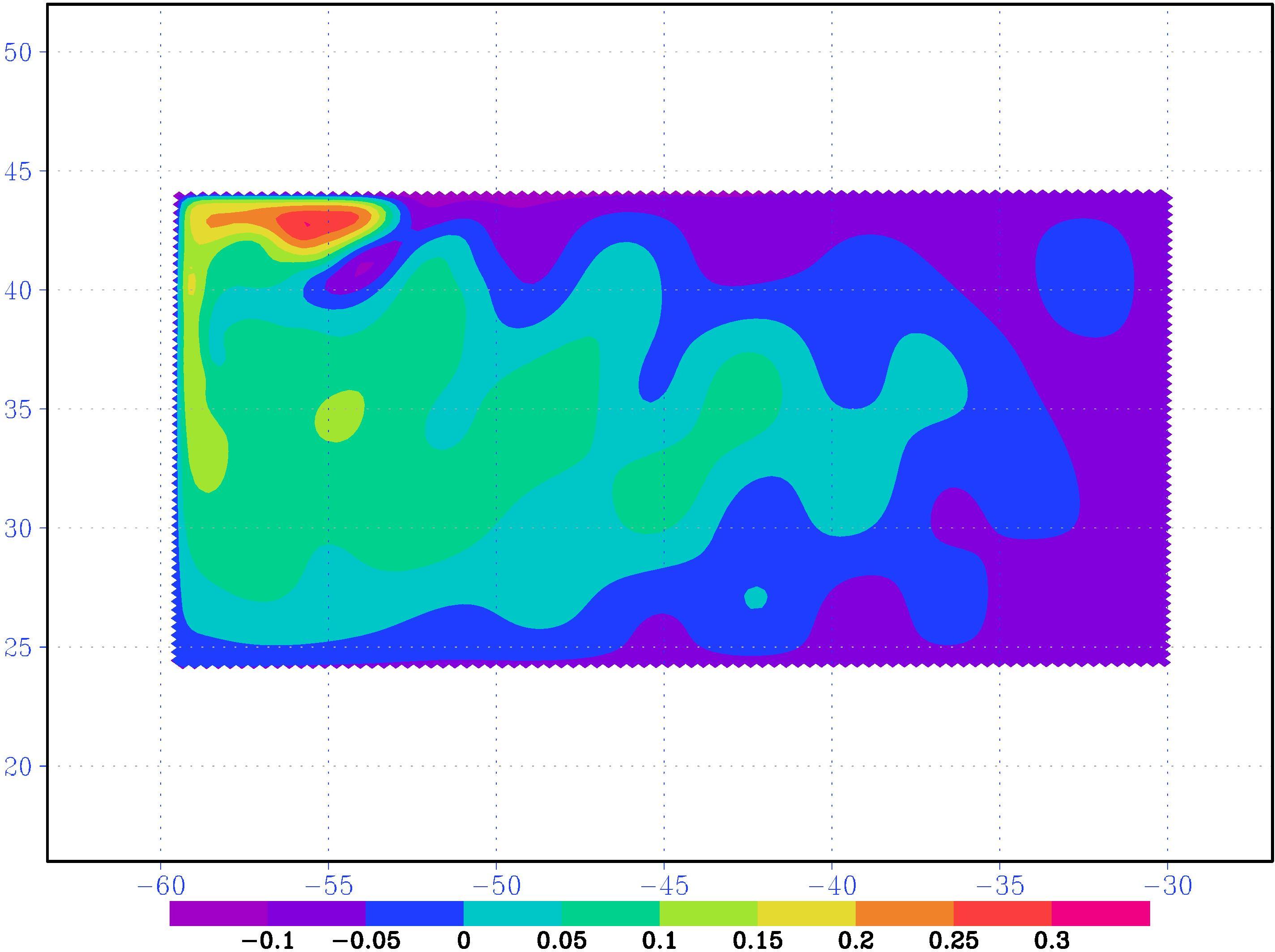}}
  \end{minipage}  
  \end{center} 
\caption{Sea surface height. Reference solution on the aligned grid (A) and the solution on the rotated grid with staircase-like boundary (B). }
 % \refstepcounter{fig}
\label{sg-ref}
\end{figure*}

At the first glance,  optimal boundary conditions are evident in this case and can be obtained without such a  costly procedure as variational data assimilation. Instead of imposing the impermeability condition on the staircase, we should apply a cut-cells approximation and impose this condition on the physical boundary  (slanted line in \rfg{45grid}) that coincides with the cut-cells approximation. In the presented case ($45^\circ$ rotated grid), impermeability must be represented as $u=v$ everywhere on the physical boundary ensuring only a tangential flux. Free-slip condition can be approximated in this case by the simple discretization  of its formulation \rf{vecbc}: 
\beqr
u(x,y)|_{bnd}&=&v(x,y)|_{bnd}=v(x+h,y-h), \nonumber \\
 v(x,y)|_{bnd}&=&u(x,y)|_{bnd}=u(x+h,y-h)
\label{faux}
\eeqr
for the part of the boundary schematically represented in the \rfg{45grid}. In other words, for $u_{i,i}$ at the  boundary we put the value of the corresponding $v_{i,j-1}$ at the  closest node in the orthogonal direction. Similarly,  the value of $v_{i,i}$ at the boundary is imposed to be equal to   $u_{i+1,i}$. Under these conditions, the value of the relative vorticity on the boundary ($\omega_{i,i}$   in \rfg{45grid}), approximated in a conventional  way,  vanishes automatically:
\beqr
\omega_{i,i} &=& \fr{v_{i,i-1}-v_{i-1,i-1}}{h}-\fr{u_{i,i}-u_{i,i-1}}{h}=\nonumber\\
 &=&\fr{v_{i,i-1}-u_{i,i-1}}{h}-\fr{v_{i,i-1}-u_{i,i-1}}{h}=0\nonumber
\eeqr

Unfortunately, the most evident idea  occurs to be   a wrong one.  Let us consider 
 linearized and  simplified part of equations \rf{1.1},\rf{1.2} that corresponds to the Coriolis parameter:
\beq
\der{u}{t}=
 \overline{\overline{v}^x}^y \overline{f}^y,\quad
 \der{v}{t}=- \overline{\overline{ u}^x}^y\overline{f}^x \nonumber
 \eeq
The matrix of this system must have only imaginary eigenvalues that ensures only oscillatory modes in the solution. 
This requirement is satisfied when no tangential flux is admitted on the boundary, but, accepting this flux, we add exponentially growing modes to the solution. Indeed,  consider for example the equation written for the $u_{2,1}$ node   in \rfg{45grid}:
$$
\der{u_{2,1}}{t}=\fr{v_{1,1}+v_{2,1}+v_{1,0}+v_{2,0}}{4}\fr{f_{2,2}+f_{2,1}}{2}
$$

 The right-hand side of this equation contains $v$ interpolated to this node. But, the value of $v_{1,1}$ on the boundary is prescribed  to be equal to 
$u_{2,1}$ by the equation \rf{faux}. That means,  the matrix of the system contains diagonal elements
$$
\der{\mathbf{u_{2,1}}}{t}=\fr{f_{2,2}+f_{2,1}}{8}\mathbf{u_{2,1}} +\cdots
$$
 and eigenvalues with non-zero real parts (as negatives and positives)  resulting in existence of   exponentially growing modes in the solution. 

Consequently, even if we know the exact  physical boundary of the domain,  straightforward approximation of the  prescribed conditions at this boundary may lead to  unstable behavior of the model. 

Dealing with real configurations of ocean models, the  exact  boundary is not known. Continental coastlines have a very complex, even fractal, structure and it is not evident how to approximate them by the model grid. 
One way to find this approximation consist in assimilating external data controlling the approximation of differential operators near the boundary. In this paper, we assimilate the data, produced by the same model on the aligned grid into the model on the rotated grid. We use the set of  three assimilation windows (15, 30, and 50 days)   in order to reduce computational time allowing 20, 50, and 20 iterations in each window respectively. As it has been noted above, both initial and boundary conditions are controlled in each window, but  in short windows, mostly the  initial state is modified, while in long windows essentially the boundary conditions are  corrected. Indeed, is we analyze the model solution on the rotated grid, we see that during first 15 days the solution is relatively noisy due to  errors committed in the interpolation of the initial point on the rotated grid, while later, the interpolation noise is already dumped  and the shortening of the boundary current becomes the principal feature. 

Discretization  of 4 differential operators $\Dx, \Dy$ near the boundary is controlled in all the experiments below: two in the lateral dissipation in the equations \rf{1.1}, \rf{1.2} and two in the calculation of the relative vorticity \rf{1.6}. Moreover, according to \rf{s-1}, discretizations  of operators ${\Dy}u$ and  ${\Dx}v$ are controlled at two points: at the boundary ($\omega_{i,i}$ points  in \rfg{45grid}) and near the boundary ($\omega_{i+1,i}$). In total, 10 control coefficients $\alpha$ are controlled near each boundary point: 3 coefficients in the $\Dx v$ expression, 3  in the $\Dy u$, 2 in $\Dx \omega$ and 2 in $\Dy \omega$.

\beqr
(\Dx v)_{i,i,k}&=& \fr{ \alpha^{Dxv}_0 v_{i,i-1,k} }{h}\nonumber\\
(\Dx v)_{i+1,i,k}&=& \fr{ \alpha^{Dxv}_1 v_{i,i-1,k}+\alpha^{Dxv}_2 v_{i+1,i-1,k} }{h}
\nonumber\\
(\Dy u)_{i,i,k}&=& \fr{ \alpha^{Dyu}_0 u_{i,i-1,k} }{h}
 \\
(\Dy u)_{i,i-1,k}&=& \fr{ \alpha^{Dyu}_1 u_{i,i-2,k}+\alpha^{Dyu}_2 u_{i,i-1,k} }{h}
\nonumber\\
(\Dx \omega)_{i,i-1,k}&=&\fr{ \alpha^{Dx\omega^l}_1 \omega_{0,j,k}+\alpha^{Du\omega^l}_2 \omega_{1,j,k} }{h}
\nonumber\\
(\Dy \omega)_{1/2,j,k}&=&\fr{ \alpha^{Dy\omega^l}_1 \omega_{0,j,k}+\alpha^{Dy\omega^l}_2 \omega_{1,j,k} }{h}
\nonumber
\eeqr

\begin{figure}[h]
  \begin{center}
  \begin{minipage}[l]{0.45\textwidth}
  \centerline{\includegraphics[angle=0,width=0.99\textwidth]{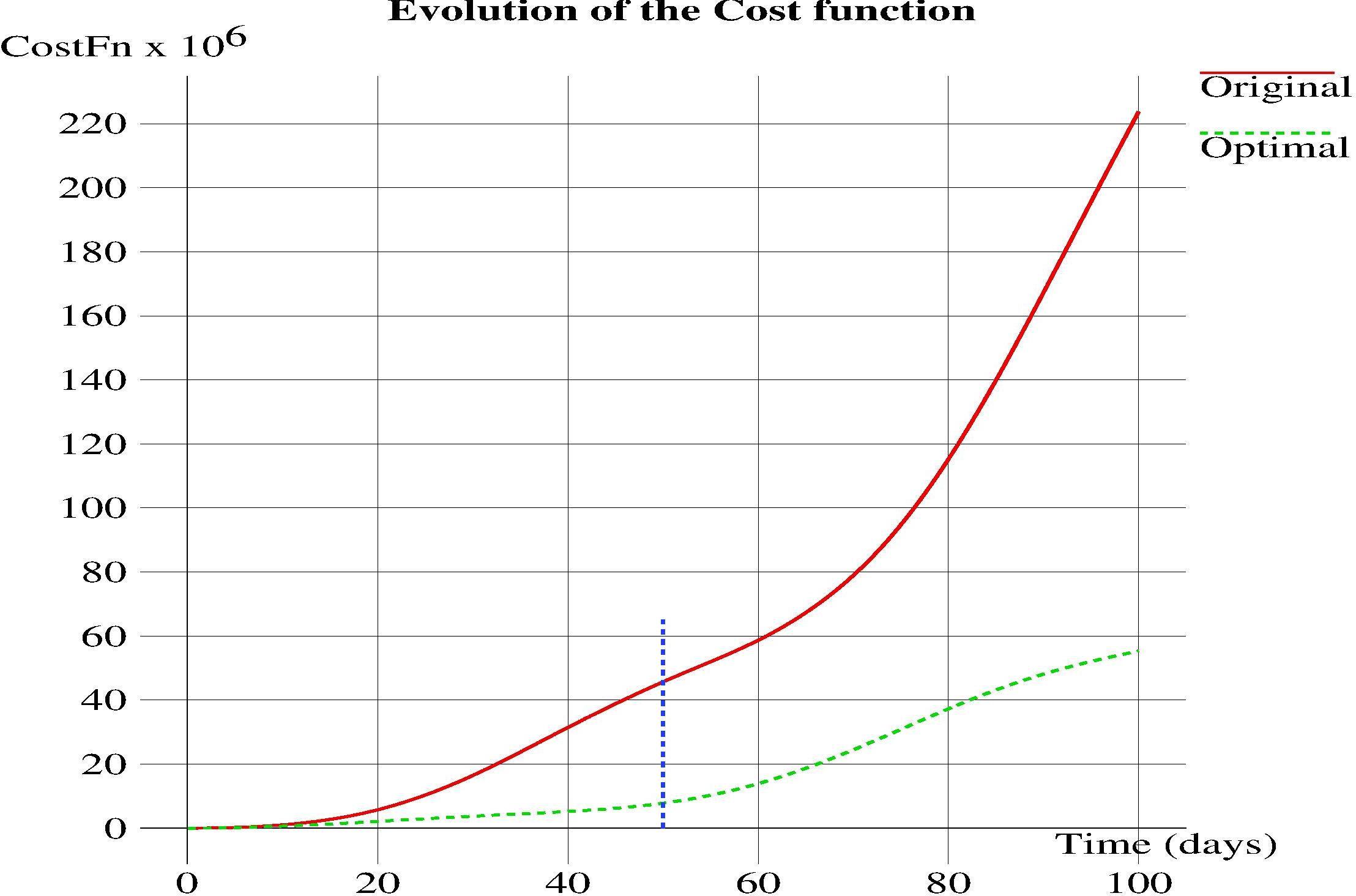}}
  \end{minipage} 
  \end{center} 
\caption{Evolution of the cost function during 100 days: Classical boundary conditions (solid red line), Optimal initial and boundary conditions (dashed green line).  }
 % \refstepcounter{fig}
\label{oib45-valj}
\end{figure}

Evolution of the cost function with conventional and optimal initial-boundary conditions is presented in \rfg{oib45-valj}. One can see that at the end of the longest  assimilation window  (50 days) the cost function value  is divided by 6 and, even 50 days later, the optimal  cost function remains 4 times lower than the original one. This fact indicates the importance of modification of the boundary conditions because the influence of the starting point becomes negligible after 100 days integration.

\begin{figure*}[t]
  \begin{center}
  \begin{minipage}[l]{0.40\textwidth}
     A. \\ 
  \centerline{\includegraphics[angle=0,width=0.99\textwidth]{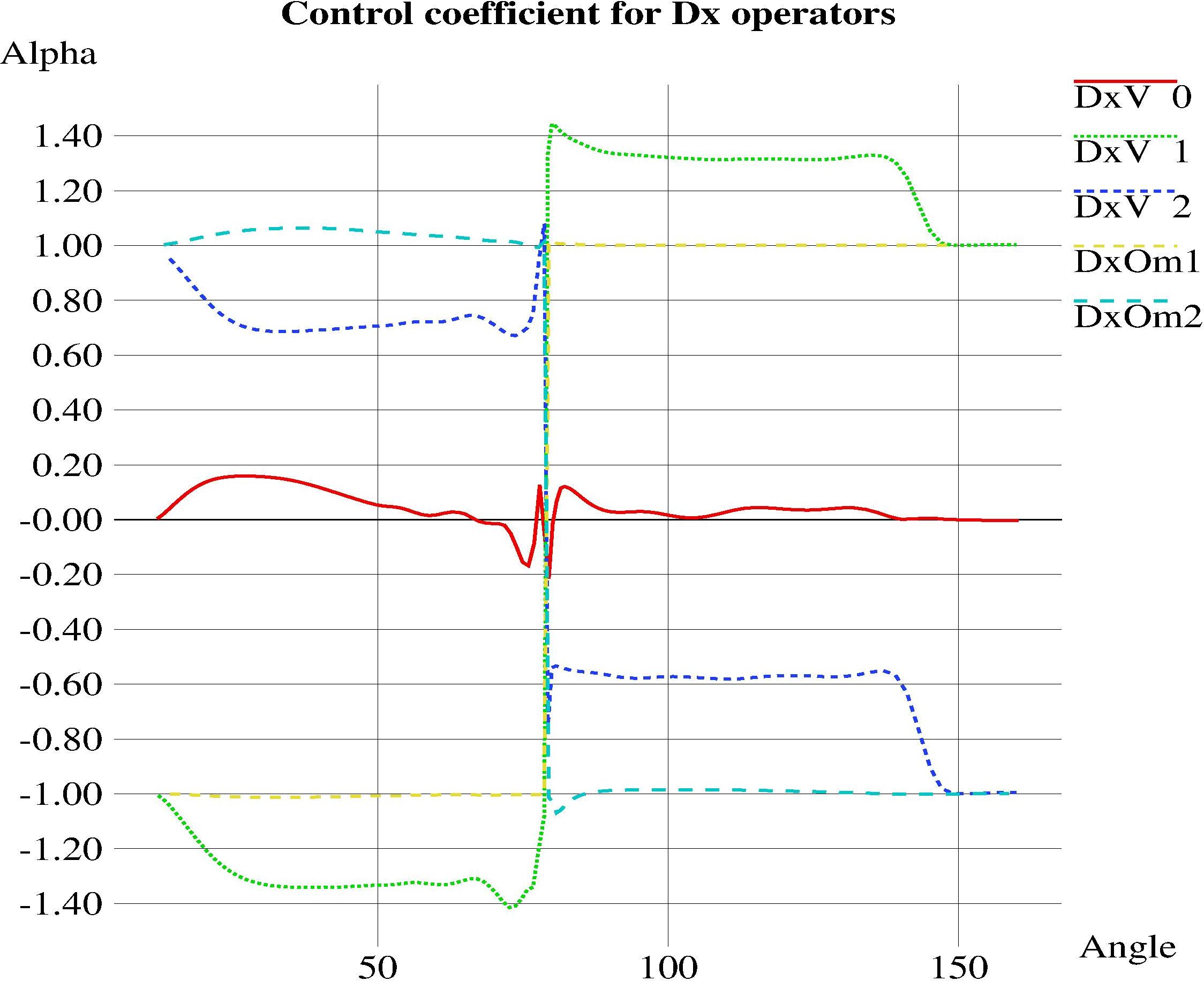}}
  \end{minipage} 
  \begin{minipage}[l]{0.40\textwidth}
    \hfill B. \\ 
  \centerline{\includegraphics[angle=0,width=0.99\textwidth]{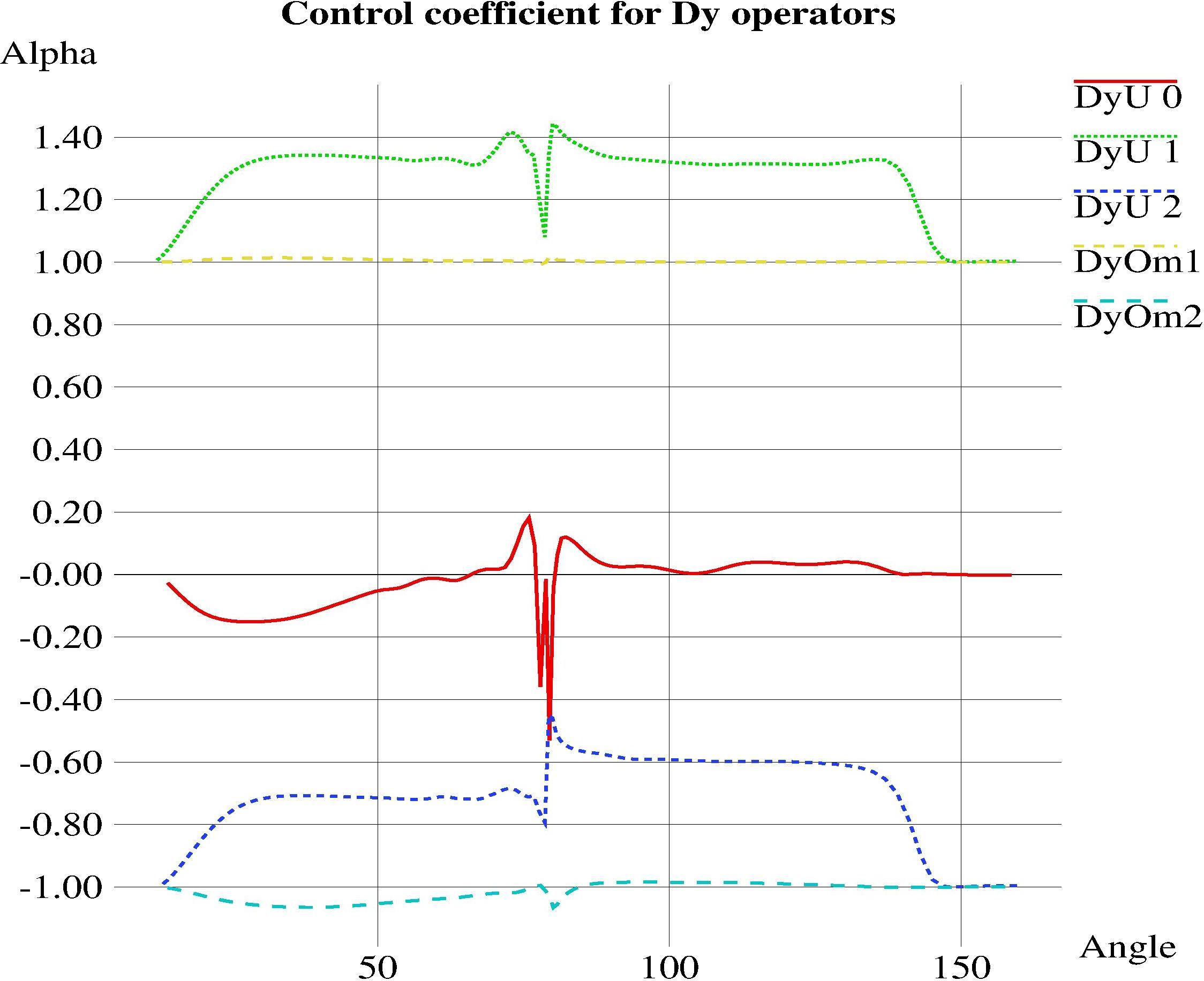}}
  \end{minipage} 
  \end{center} 
\caption{Optimal coefficients $\alpha$ for the Western and the Northern boundaries on the $45^\circ$ rotated grid for derivatives in the  $x$ direction (A) and in the $y$ direction (B).  }
 % \refstepcounter{fig}
\label{oib45xy}
\end{figure*}

Optimal coefficients $\alpha$ for the Western and the Northern boundaries are shown in \rfg{oib45xy}. The horizontal axis points out  the position  on the boundary indicating the  angle with  the south-western direction in the latitude-longitude coordinates. The angle $12^\circ$ corresponds approximately to the lower left corner of the rectangle,  $78^\circ$ --- to the upper left corner and $198^\circ$ --- to the upper right corner.  Coefficients are plotted, however, up to $160^\circ$ only, because the control is negligible beyond  the Northern boundary current (about $145^\circ$,  see \rfg{sg-ref}).  No significant modification of $\alpha$ is observed near the Eastern and Southern boundaries as well. This can be explained by the absence of the boundary flow at these places. So far, there is no current, the flow is already close to the reference one and the  data assimilation does not need to modify anything.  

Coefficients $\alpha$ for $\Dx$ operators change sign at the upper left corner ($78^\circ$) because approximations are calculated by different formula: \rf{s-1} on the left and  \rf{rightbnd} on the right from $78^\circ$.

In this figure one can see significant differences of optimal coefficients with respect to the classical derivatives approximations for operators $\Dx v, \Dy u$ that are used in the vorticity calculation. 
The strongest modification is observed in the approximation of the vorticity at adjacent to boundary nodes ($\omega_{i+1,i}$ in \rfg{45grid}). At these nodes, optimal vorticity is approximated with the coefficients 
$\alpha^{Dxv}_1, \alpha^{Dxv}_2$ and $ \alpha^{Dyu}_1, \alpha^{Dyu}_2$ which values may differ from the conventional $+1$ or $-1$ by $\pm 0.4$ (dashed green and blue lines in \rfg{oib45xy}).

In order to see the influence of the optimal discretization of derivatives on the model solution on long-time scales, we run the model for 800 days staring from optimal initial point found in the data assimilation  and plot the solution obtained at the end of this run in \rfg{sg-opt}A. Despite the largest assimilation window was as short as 50 days, the sea surface height of the optimal solution after 800 days model run is quite close to the reference one shown in \rfg{sg-ref}A. The major  difference between optimal and reference flows consists in a slightly overestimated wave activity. 

 Analyzing the magnitude of the control applied to different $\alpha$, we can see that  approximations of the dissipation operators in equations \rf{1.1}, \rf{1.2} have been modified only a little (by less than 0.05) (yellow and cyan dashed lines in \rfg{oib45xy}).  Optimal vorticity value at the boundary, calculated with $\alpha^{Dxv}_0, \alpha^{Dyu}_0$, remains also close to 0 all along the boundary except in the lower left corner (solid red lines in \rfg{oib45xy}). These modifications  seem to be insignificant and
we can suppose that optimal vorticity calculation near the boundary is sufficient to compensate the effect of the staircase shaped boundary. 

Following this hypothesis,  we substitute for $\alpha^{Dxv}$, $ \alpha^{Dyu}$ their "typical" optimal values ($\pm 1.4, \pm 0.6$ at the adjacent nodes). The vorticity approximation near the Western coast becomes:
\beqr
\omega_{i+1,i} &=& \fr{ 0.6 v_{i+1,i-1} -1.4 v_{i,i-1}}{h}-\fr{ 1.4 u_{i+1,i} -0.6 u_{i+1,i-1}}{h}
\nonumber\\
&=&\fr{ v_{i+1,i-1} - v_{i,i-1}}{h}-\fr{  u_{i+1,i} - u_{i+1,i-1}}{h} - 
\nonumber\\
&-&
\fr{0.8}{h} \fr{v_{i+1,i-1} + v_{i,i-1}+  u_{i+1,i} + u_{i+1,i-1}}{2}\nonumber\\
&=&
\delta_x\left[v\right]-\delta_y\left[u\right] - \fr{0.8}{h}(\overline{u}+\overline{v})
\label{mdf}
\eeqr
Similar modification is also made for the vorticity calculation at nodes adjacent to the Northern boundary:
\beqr
\omega_{i,j} &=& \fr{ 1.4 v_{i,j-1} -0.6 v_{i-1,j-1}}{h}-\fr{ 1.4 u_{i,j} -0.6 u_{i,j-1}}{h}
\nonumber\\
&=&
\delta_x\left[v\right]-\delta_y\left[u\right] - \fr{0.8}{h}(\overline{u}-\overline{v})
\label{mdfnorth}
\eeqr
  Moreover, we suppose also that the vorticity may be modified near the whole boundary, and namely near the Eastern and the Southern coasts despite these modifications are not required by  the data assimilation. To verify this hypothesis, we perform an additional model run from the optimal initial point with the only modification  defined by  equations \rf{mdf}, \rf{mdfnorth} but applied to the whole boundary. The SSH obtained on the 800th day in this run can be seen in \rfg{sg-ref}B. Indeed, two patterns in \rfg{sg-ref} are similar, that confirms the hypothesis: optimization  is necessary for the vorticity approximation only.

Of course, the estimation of these "typical" optimal values is only qualitative. No sensitivity analysis of these optimal coefficients is made in this paper and no confidence interval is determined because we address rather the existence of the optimal discretization than precise estimates of coefficients. 
  
The equation \rf{mdf} shows that  $-0.8(\overline{u}+\overline{v})/h$ has been added to the classical approximation of the relative vorticity at adjacent nodes. So far, due to free-slip condition, the tangential velocity $u+v$ does not vanish,  we have got minus first order of the vorticity approximation. 
Similar situation is observed near the northern boundary: corresponding value  ($-0.8(\overline{u}\mathbf{-}\overline{v})/h$ in this case) is also added to the conventional approximation.

\begin{figure*}[t]
  \begin{center}
  \begin{minipage}[l]{0.48\textwidth}
   A. \\ 
  \centerline{\includegraphics[angle=0,width=0.99\textwidth]{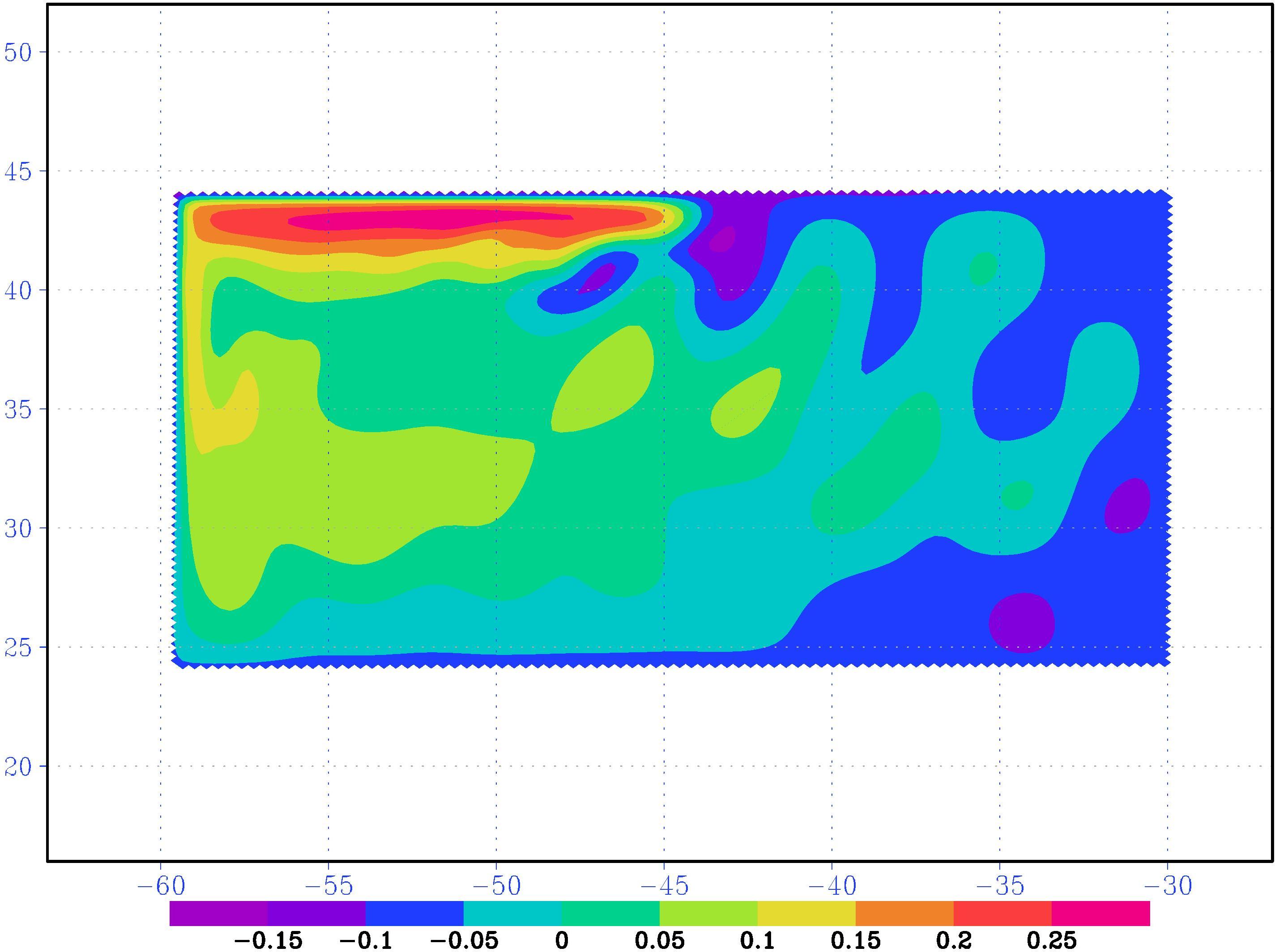}}
  \end{minipage} 
  \begin{minipage}[r]{0.48\textwidth} 
      \hfill B. 
  \centerline{\includegraphics[angle=0,width=0.99\textwidth]{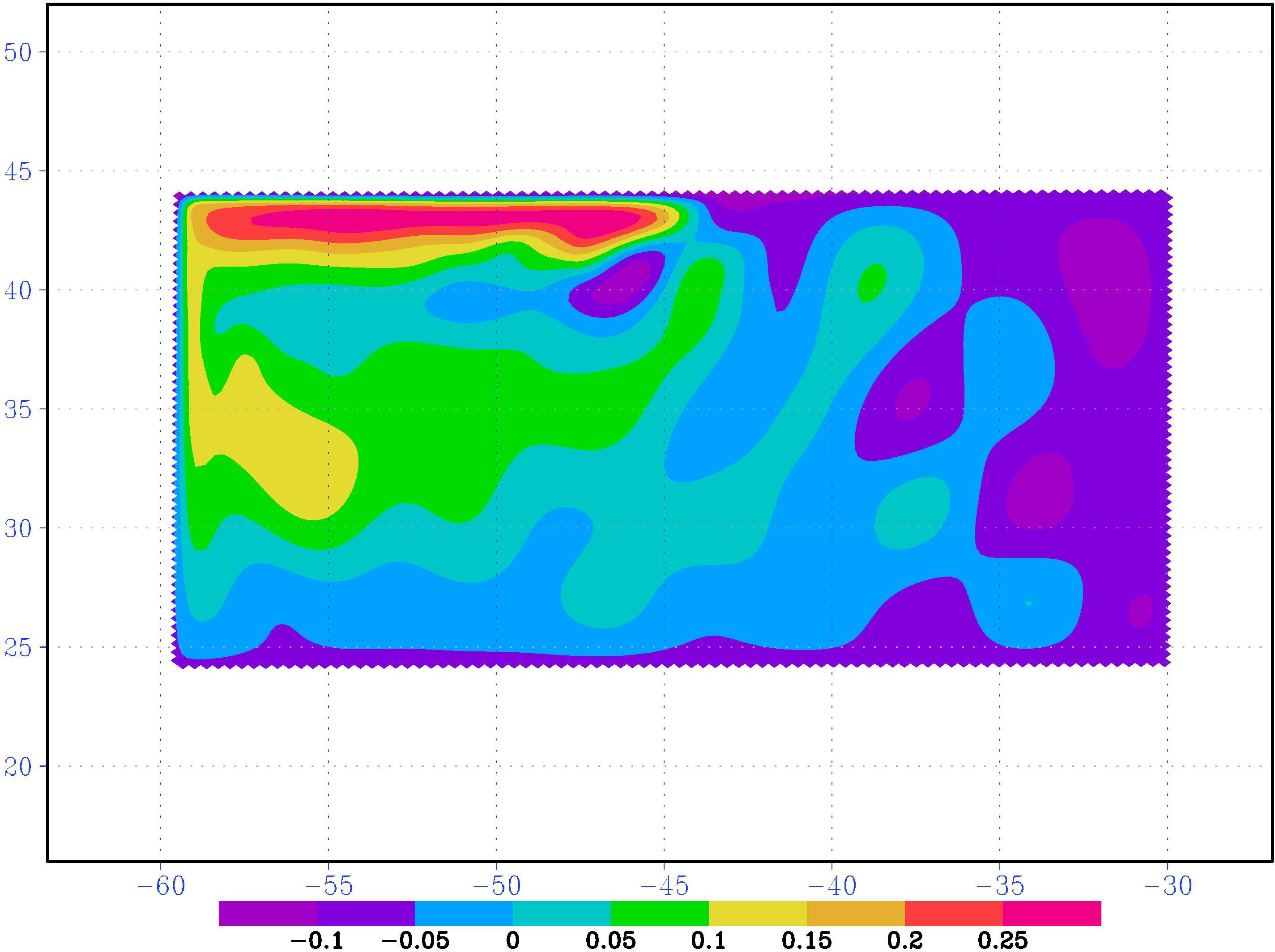}}
  \end{minipage}  
  \end{center} 
\caption{Sea surface height on the 800th day. Solutions on the rotated grid: optimal (A)  and modified  (B) boundary conditions. }
 % \refstepcounter{fig}
\label{sg-opt}
\end{figure*}

In order to interpret the addition  of $-0.8(\overline{u}\pm\overline{v})/h$  to the approximation of the relative vorticity, we can refer to \cite{BlayoMarSys, VerronBlayo} who show the free-slip condition must be formulated as $\omega_{{bnd}}=\fr{\vec{U}\cdot\vec{\tau}}{R}$ instead of $\omega_{{bnd}}=0$ in the case of a curvilinear boundary. Taking into account that  tangential velocity components are equal to $(\vec{U}\cdot\vec{\tau}) =(u+v)/\sqrt{2}$ near the Western coast and to $(u-v)/\sqrt{2}$ near the Northern one, the additional term in \rf{mdf}, \rf{mdfnorth}  is always proportional to the tangential velocity component 
\beq
-\fr{0.8(\overline{u}\pm\overline{v})}{h}=-\fr{ 0.8 \sqrt{2} (\vec{U}\cdot\vec{\tau})}{h}= \fr{ \vec{U}\cdot\vec{\tau}}{R}
\eeq
where $R=-\fr{h}{0.8 \sqrt{2}}$. 

Consequently, the optimal boundary determined by data assimilation  in this case  is neither a straight line, nor a staircase, but a curvilinear boundary with constant curvature radius $R=-\fr{h}{0.8 \sqrt{2}}$.
The additional term in \rf{mdf}, \rf{mdfnorth}   can be interpreted as the influence of a curvilinear boundary with curvature radius $R$ on the  vorticity: at the nearest to boundary points, the vorticity is calculated by the usual formula $\delta_x\left[v\right]-\delta_y\left[u\right]$ because this node is in ocean, but an additional term $(\vec{U}\cdot\vec{\tau})/R $ is added due to curvilinearity requirement.

So far,    the optimal curvature depends on the grid step $h$ and the resolution may influence the optimal radius $R$, we perform two additional experiments: with the double resolution ($h=1^\circ/8$) and the half resolution ($h=1^\circ/2$) keeping all other parameters (including lateral dissipation) unchanged. As well as abobe, the final state of the 1000 days spin-up is interpolated to the rotated grid  and the model runs for  800 days from this state either with the classical or with the optimal  (defined by \rf{mdf}, \rf{mdfnorth}) discretization of derivatives  in the vorticity operator. Final states of the double resolution model are shown in \rfg{doubres} and of the half-resolution model in \rfg{halfres}.

\begin{figure*}[t]
  \begin{center}
  \begin{minipage}[l]{0.48\textwidth}
   A. \\ 
  \centerline{\includegraphics[angle=0,width=0.99\textwidth]{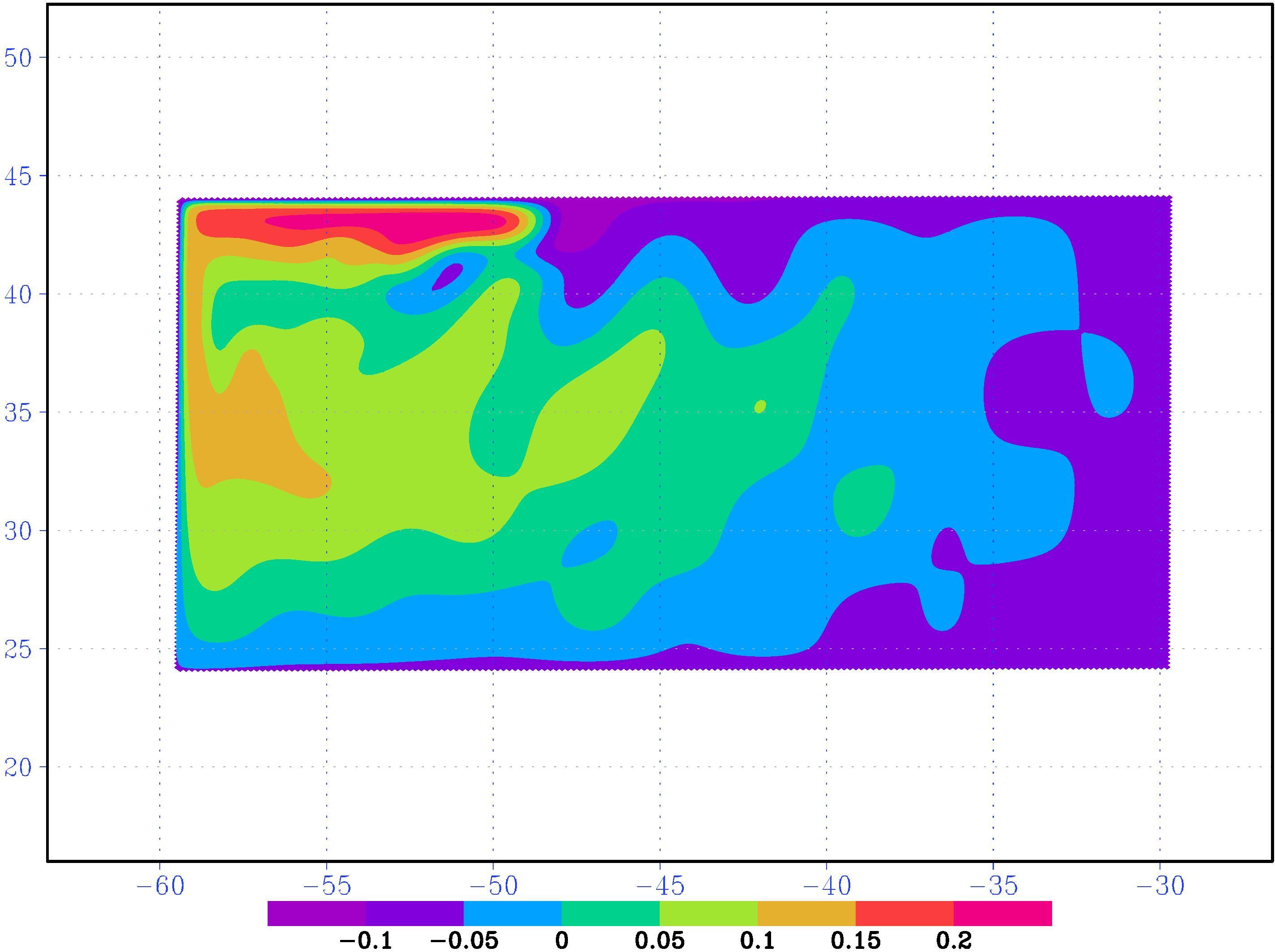}}
  \end{minipage} 
  \begin{minipage}[r]{0.48\textwidth} 
      \hfill B. 
  \centerline{\includegraphics[angle=0,width=0.99\textwidth]{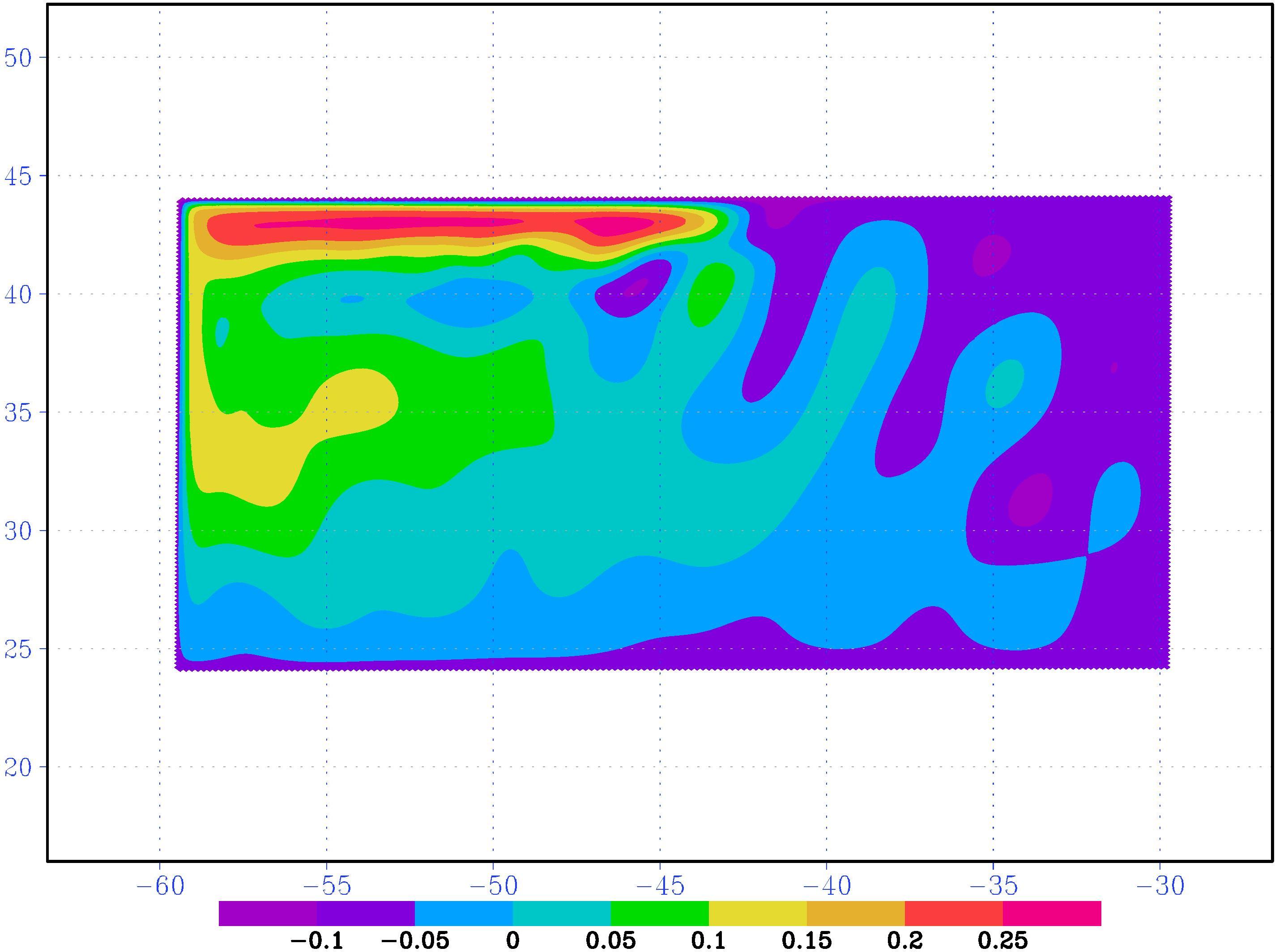}}
  \end{minipage}  
  \end{center} 
\caption{Sea surface height on the 800th day for the ($h=\fr{1}{8}^\circ$) resolution. Solutions on the rotated grid: classical (A)  and modified by \rf{mdf}  (B) boundary conditions. }
 % \refstepcounter{fig}
\label{doubres}
\end{figure*}

\begin{figure*}[t]
  \begin{center}
  \begin{minipage}[l]{0.48\textwidth}
   A. \\ 
  \centerline{\includegraphics[angle=0,width=0.99\textwidth]{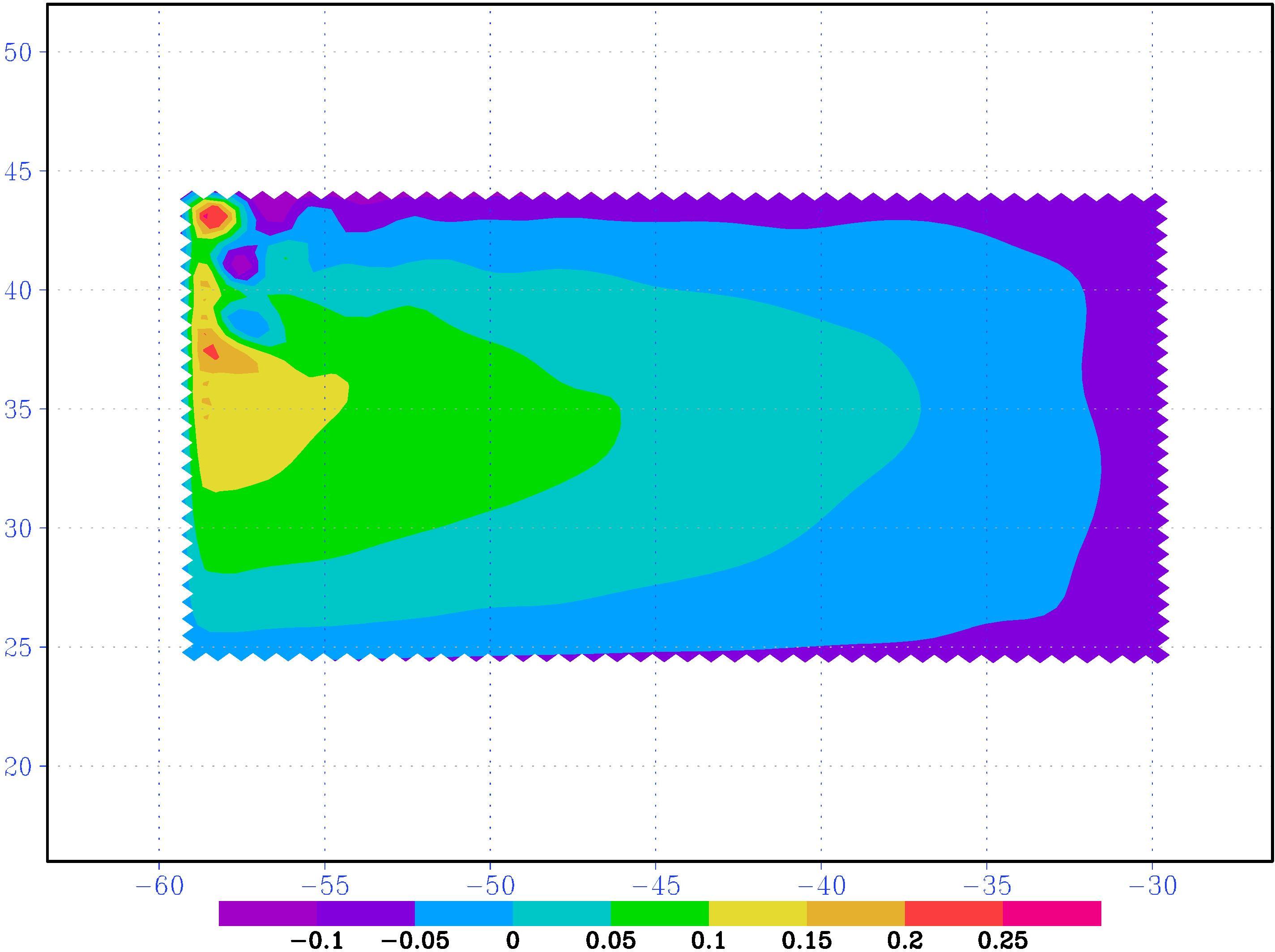}}
  \end{minipage} 
  \begin{minipage}[r]{0.48\textwidth} 
      \hfill B. 
  \centerline{\includegraphics[angle=0,width=0.99\textwidth]{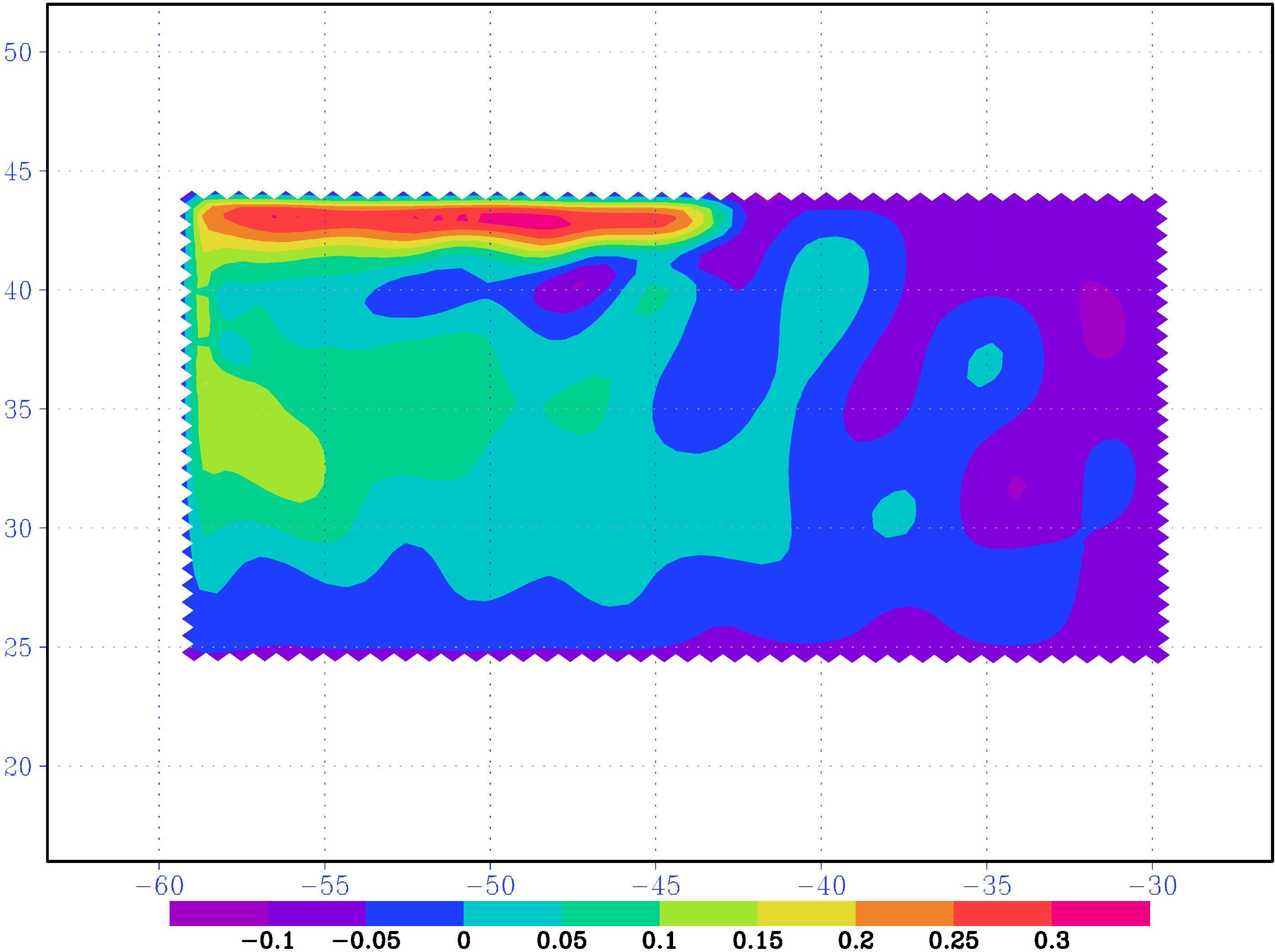}}
  \end{minipage}  
  \end{center} 
\caption{Sea surface height on the 800th day for the ($h=\fr{1}{2}^\circ$) resolution. Solutions on the rotated grid: classical (A)  and modified by \rf{mdf}  (B) boundary conditions. }
 % \refstepcounter{fig}
\label{halfres}
\end{figure*}

One can see in these figures that finer resolution  improve the solution with the classical approximation of the boundary conditions: the northern boundary current is longer in \rfg{doubres}A than in \rfg{sg-ref}B while this current  is almost absent \rfg{halfres}A. On the other hand, the model with optimal discretization of the vorticity provides quite similar results independently on the reslution: \rfg{halfres}B, \rfg{sg-opt}B, \rfg{doubres}B. 
 
\section{Single gyre forcing, $30^\circ$ rotation}

More complicated case  of the rotation on $30^\circ$ is discussed in this section. We consider the model with all the same parameters as described above on the $h=\fr{1}{4}^\circ$ resolution grid but the staircase is not uniform. The same data assimilation experiments have been performed. 

The  sea surface height on the 800th day integration of the model on the $30^\circ$ rotated grid is shown in \rfg{30deg}A. One can see, a non-uniform staircase-like boundary degrades more the solution than the uniform staircase: the Northern boundary current is  shorter and its amplitude is smaller than in \rfg{sg-ref}B.  But, as well as in the case of $45^\circ$ rotated grid,  optimal condition allows to get a similar SSH pattern  as the reference one \rfg{sg-ref}A.

\begin{figure*}[t]
  \begin{center}
  \begin{minipage}[l]{0.48\textwidth}
   A. \\ 
  \centerline{\includegraphics[angle=0,width=0.99\textwidth]{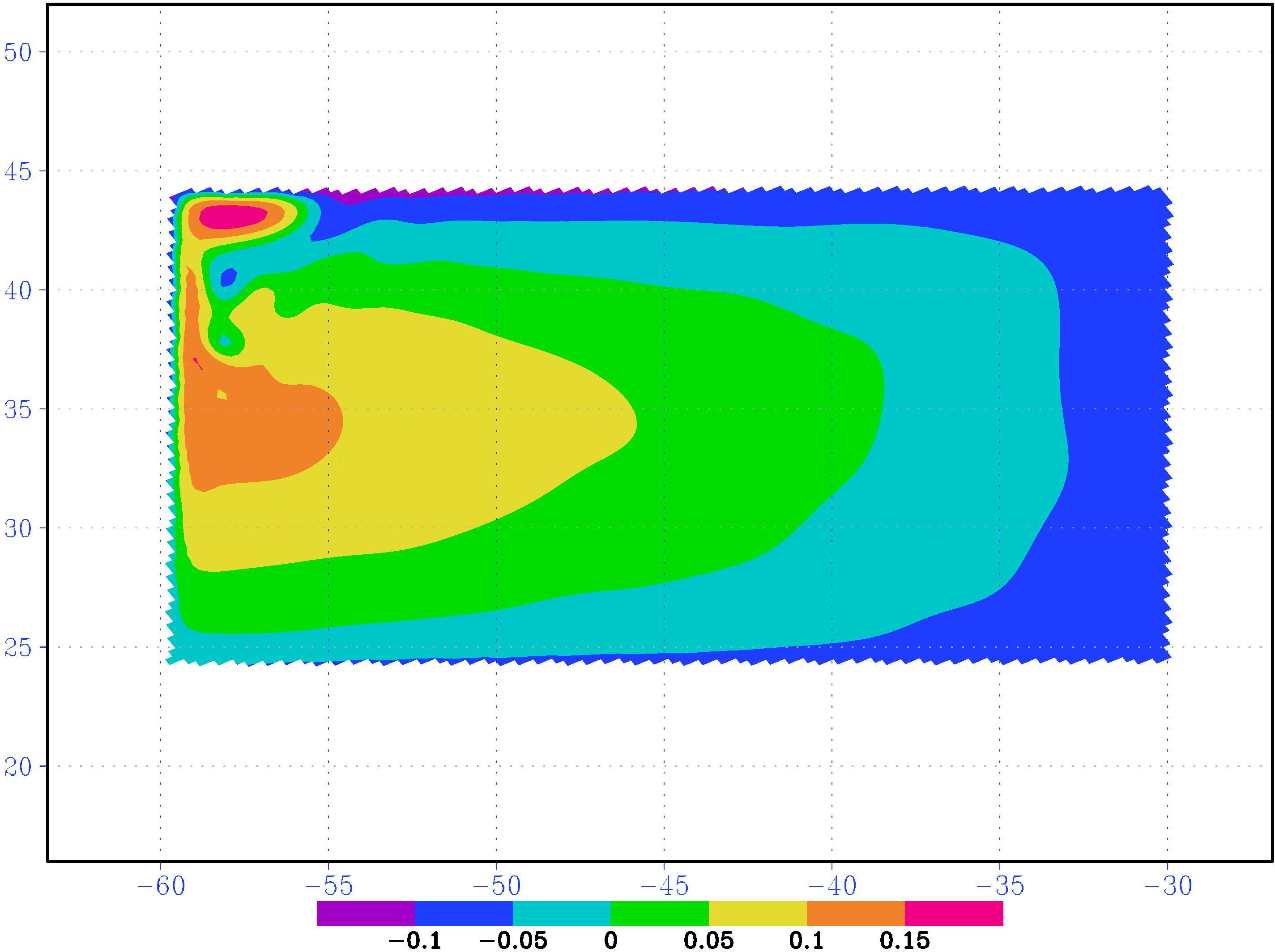}}
  \end{minipage} 
  \begin{minipage}[r]{0.48\textwidth} 
      \hfill B. 
  \centerline{\includegraphics[angle=0,width=0.99\textwidth]{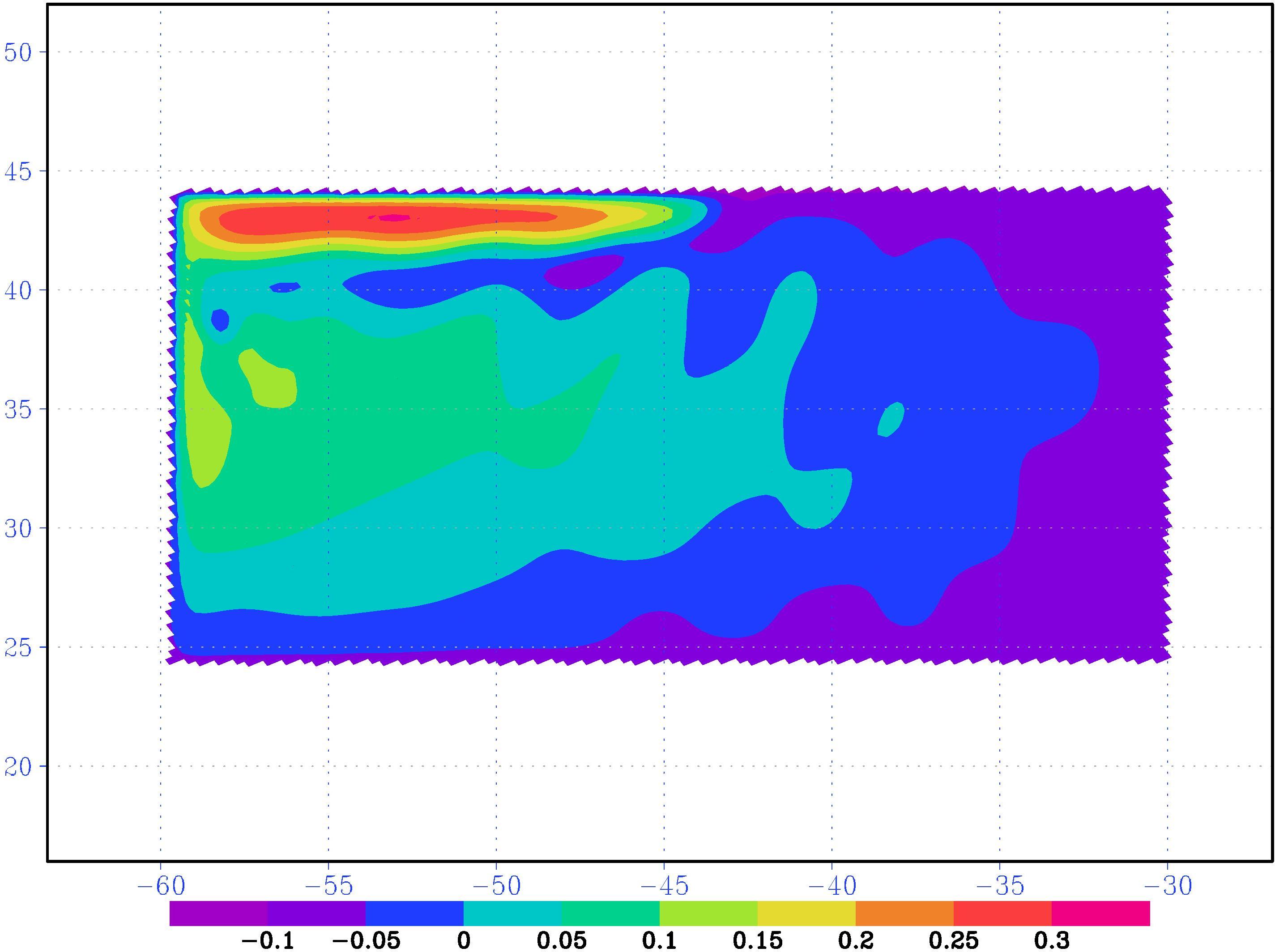}}
  \end{minipage}  
  \end{center} 
\caption{Sea surface height on the 800th day. Solutions on the $30^\circ$ rotated grid: classical (A)  and optimal  (B) boundary conditions. }
 % \refstepcounter{fig}
\label{30deg}
\end{figure*}

Optimal coefficients $\alpha$ at the Western and the Northern boundaries are plotted in \rfg{oib30xy}. The horizontal axis  indicates   the  angle with the   direction from the center to the lower-left corner of the rectangle. The  upper left corner corresponds approximately to the angle $64^\circ$ and the upper right corner -- to $180^\circ$.  However, we plot the coefficients $\alpha$  up to $140^\circ$  because beyond this angle there is no flux in the reference solution and, consequently, coefficients $\alpha$ are not modified by  data assimilation. This is also true for the Eastern and Southern boundaries where all $\alpha$ are very close to their default values \rf{alpha-initguess}. 

Comparing \rfg{oib30xy} and  \rfg{oib45xy}, we can see significant differences. First of all, due to non-uniformity of the staircase, there are significant variations of  $\alpha$  from point to point. That means the 
curvature radius is no longer  constant, but specific for each particular stair varying from $R=-h$ to $R=+5h$. 
 Second,  the amplitude of variations exceeds sometimes $\pm 0.6$  while the coefficients in the previous experiment were modified by $\pm 0.4$ at maximum.  And third, contrary to \rfg{oib45xy},  coefficients $\alpha_0$  (solid red lines) used in calculation of the vorticity at the boundary are also strongly modified. Despite  the  default is $\alpha_0=0$, their values may exceed $\pm 1$.
 
 Schematic representation of optimal boundaries on $30^\circ$ and $45^\circ$ rotated grids is shown in \rfg{curvbnd}. Optimal coastline is shown as a sequence of the segments with the constant curvature $R=h\sqrt{2}$ in the case of $45^\circ$ rotated grid and with variable curvature in the case of $30^\circ$ rotated grid. Particular curvature radius near each vorticity node corresponds to particular set of coefficients $\alpha$ that is considered to be optimal to approximate the vorticity at this node. 

\begin{figure}[!h]
  \begin{center}
  \centerline{\includegraphics[angle=0,width=0.33\textwidth]{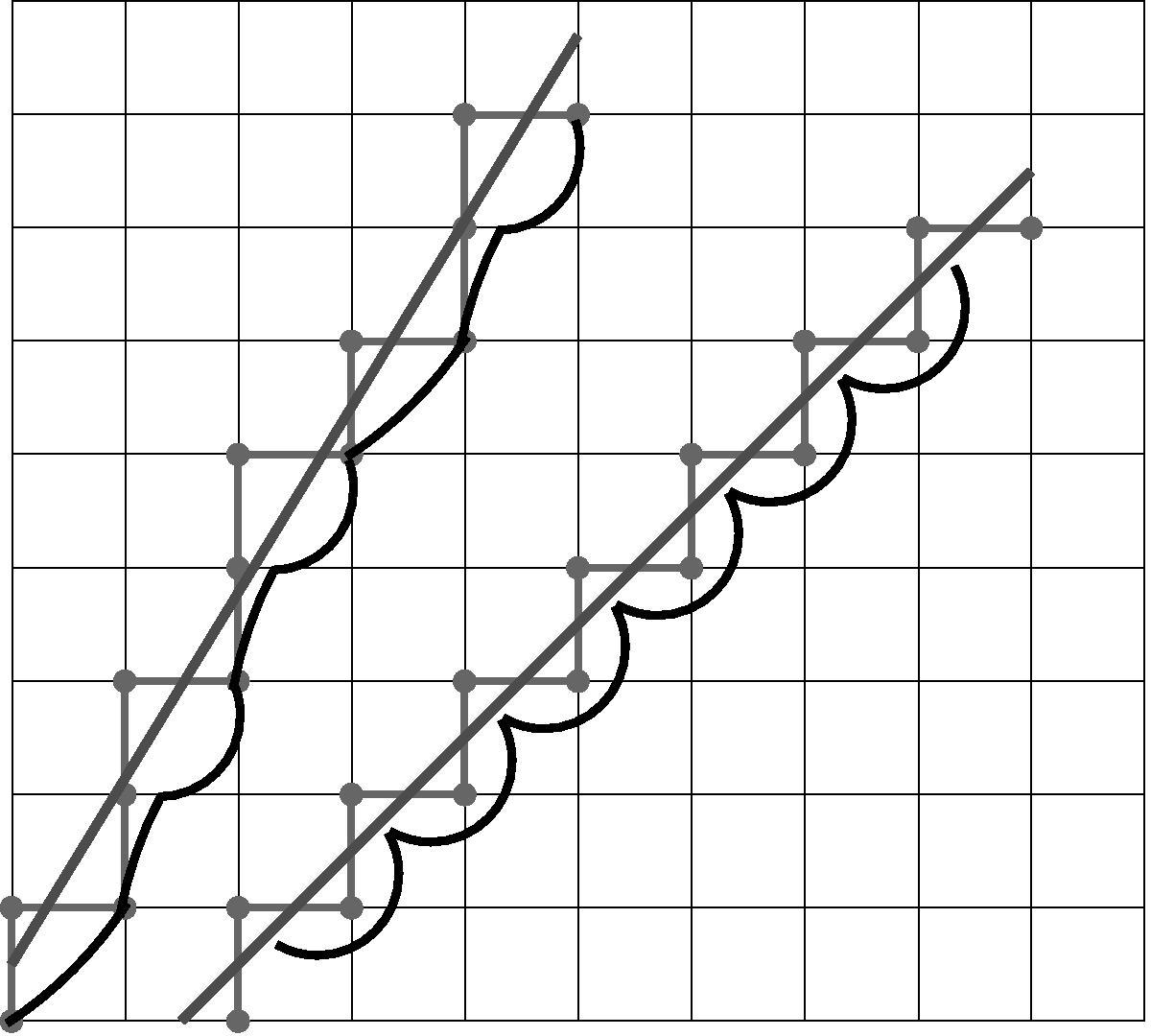}}
  \end{center} 
\caption{Schematic representation of optimal boundaries on $30^\circ$ and $45^\circ$ rotated grids: slanted grey lines represent the presumed exact coastline, black segments --- optimal coastline.   }
 % \refstepcounter{fig}
\label{curvbnd}
\end{figure}

\begin{figure*}[t]
  \begin{center}
  \begin{minipage}[l]{0.40\textwidth}
     A. \\ 
  \centerline{\includegraphics[angle=0,width=0.99\textwidth]{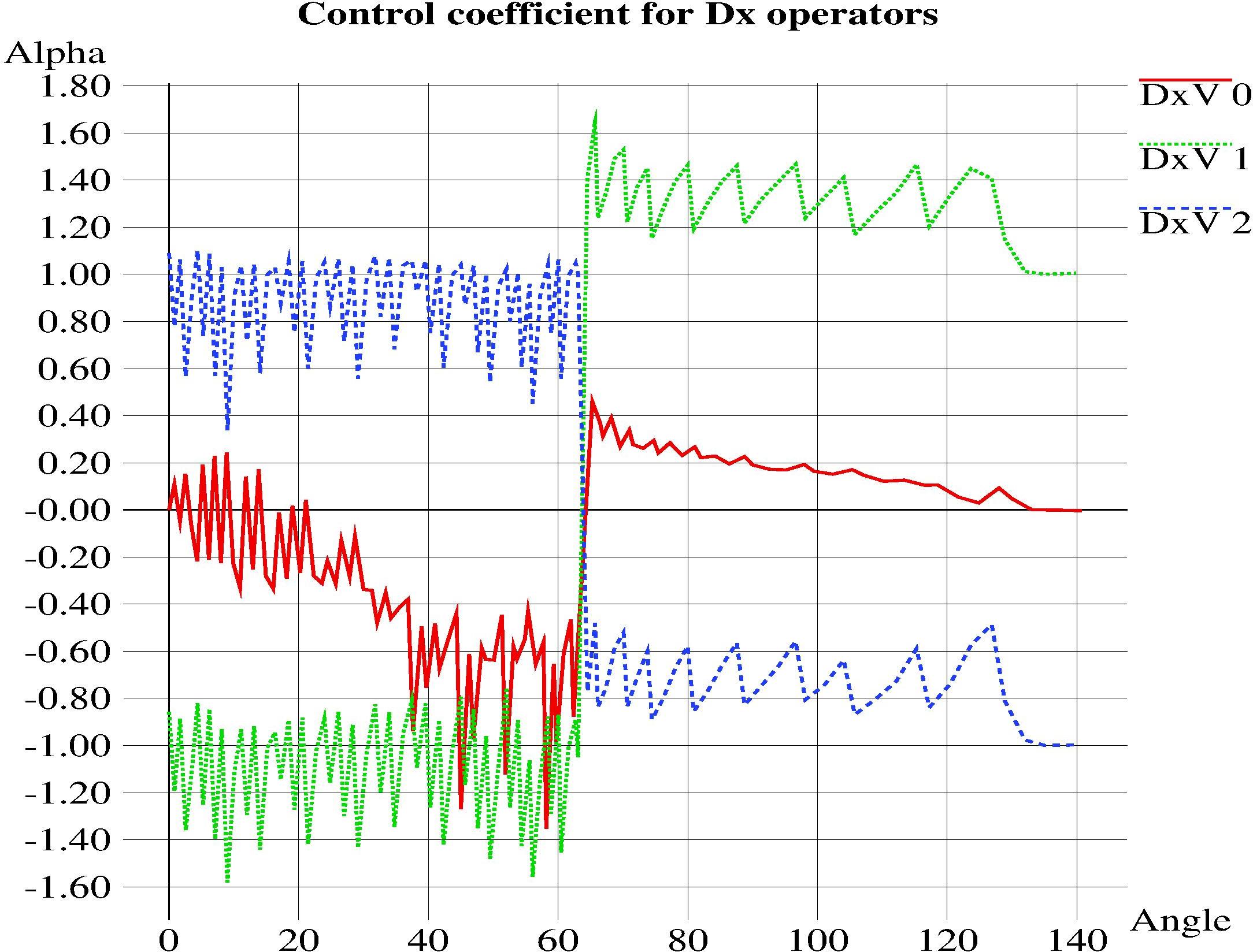}}
  \end{minipage} 
  \begin{minipage}[l]{0.40\textwidth}
    \hfill B. \\ 
  \centerline{\includegraphics[angle=0,width=0.99\textwidth]{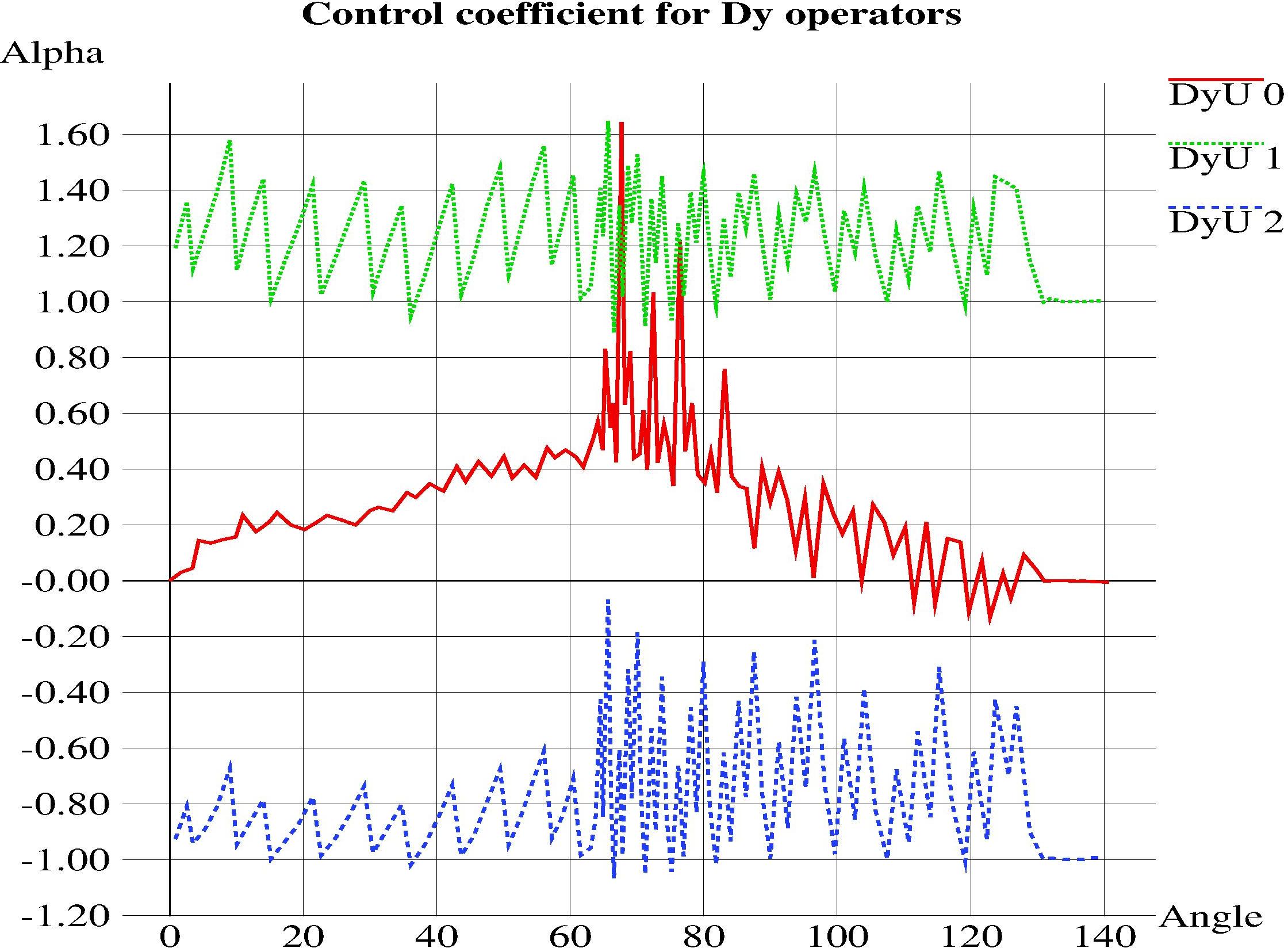}}
  \end{minipage} 
  \end{center} 
\caption{Optimal coefficients $\alpha$ for the Western and the Northern boundaries of the $30^\circ$ rotated grid for derivatives in $x$ direction (A) and in the $y$ direction (B).  }
 % \refstepcounter{fig}
\label{oib30xy}
\end{figure*}

\section{Double gyre forcing, $30^\circ$ rotation}

\begin{figure*}[t]
  \begin{center}
  \begin{minipage}[l]{0.32\textwidth}
   A. \\ 
  \centerline{\includegraphics[angle=0,width=0.99\textwidth]{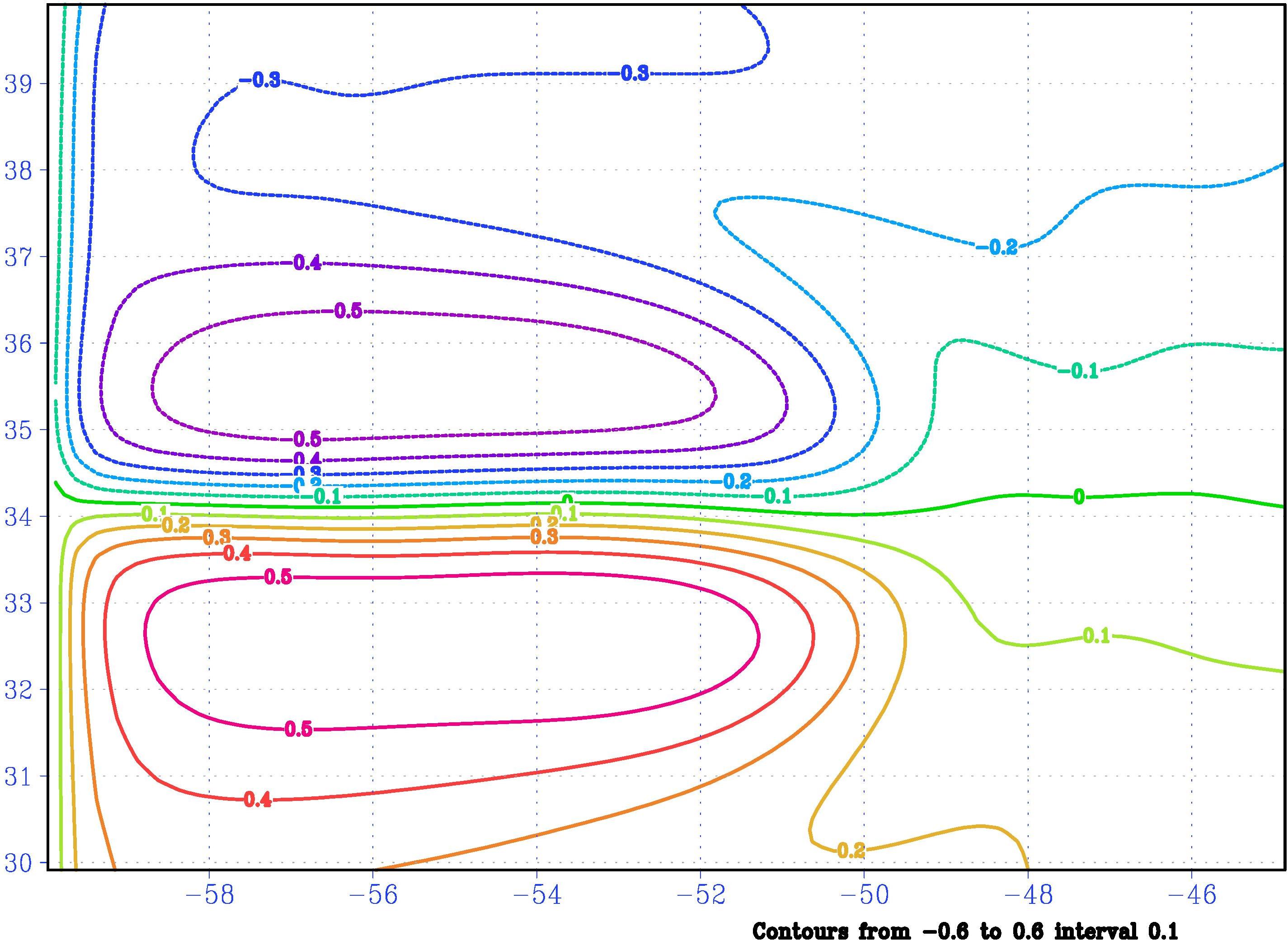}}
  \end{minipage} 
  \begin{minipage}[r]{0.32\textwidth} 
  \centerline{ B.} 
  \centerline{\includegraphics[angle=0,width=0.99\textwidth]{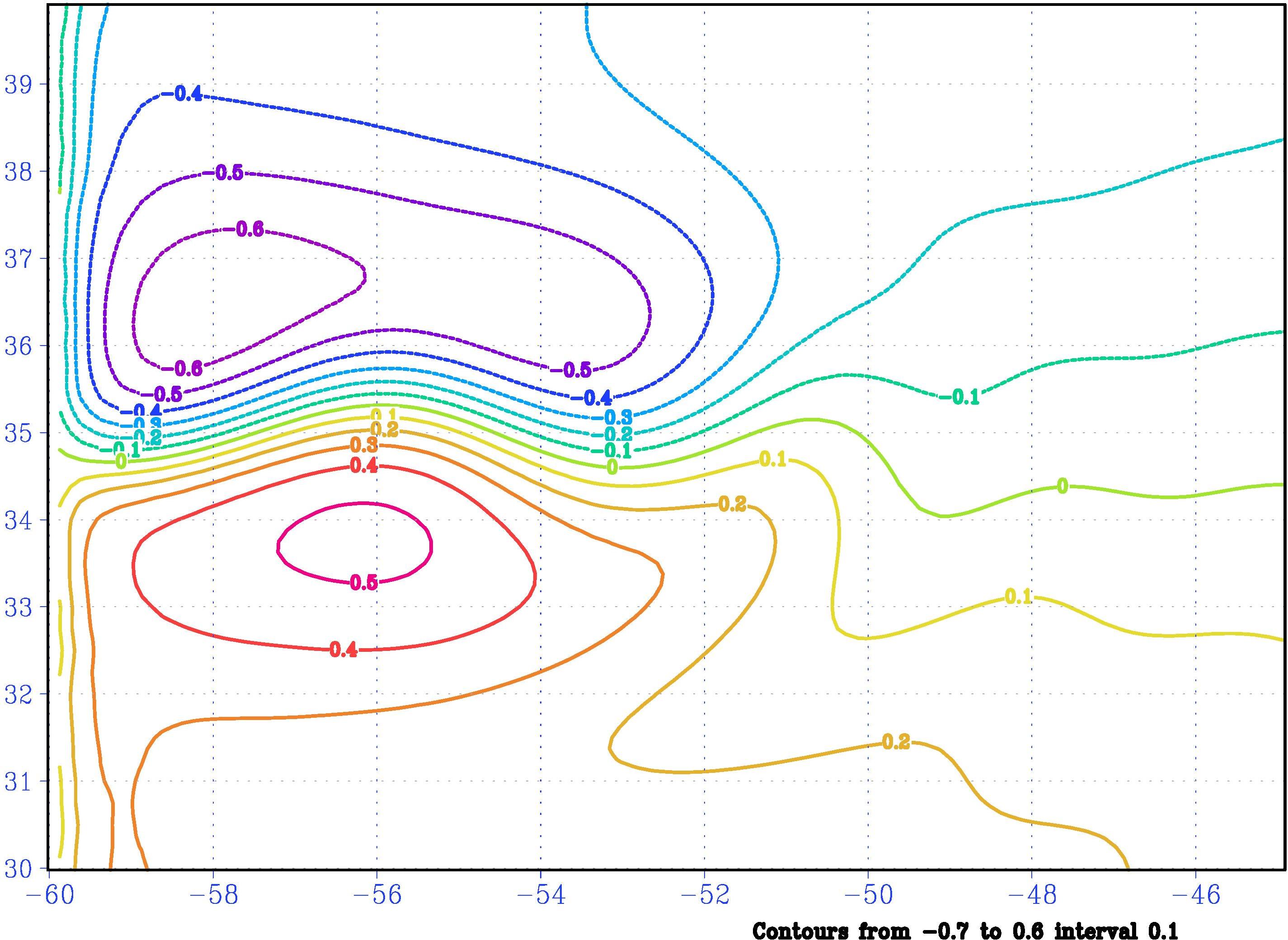}}
  \end{minipage}  
    \begin{minipage}[r]{0.32\textwidth} 
      \hfill C. 
  \centerline{\includegraphics[angle=0,width=0.99\textwidth]{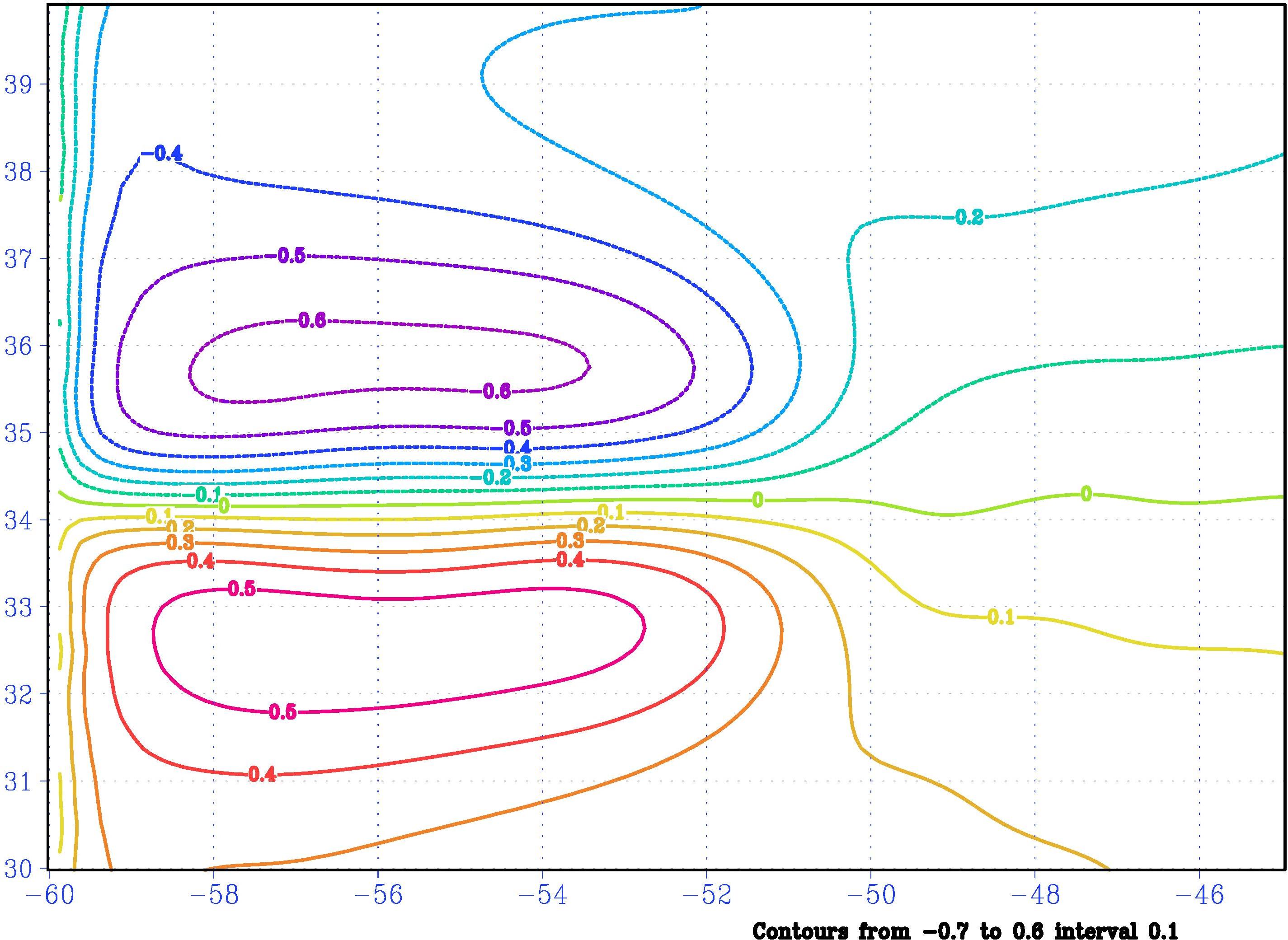}}
  \end{minipage}  
  \end{center} 
\caption{30 years average of the sea surface height. Reference model on the aligned grid (A), Model on the $30^\circ$ rotated grid: classical (B)  and optimal  (C) boundary conditions. }
 % \refstepcounter{fig}
\label{2gyre}
\end{figure*}

And finally, we discuss the most complex case in this paper: the model on the  $30^\circ$ rotated grid is subjected to the double gyre forcing ($n=2$ in \rf{forc}). Principal difference with the previous cases consists in a more complex model behavior that seems to be chaotic. The jet-stream in the middle of the rectangle exhibits irregular variations and displacements requiring to consider statistically stables long-term average fields instead of instantaneous patterns. 

The principal influence of the staircase-like approximation of the boundary consists in the modification of the direction of the jet that becomes directed slightly to the North and the displacement of the starting point of the jet for 60-70 km to the North also. These modifications are related to the Gulf Stream separation problem that attract much attention during  last 25 years (see, for example \cite{VerronBlayo, chassignet}). Supposing that boundary conditions control may also be useful in solving this problem, we perform the same assimilation experiments as described above: a sequence of three assimilation windows 15, 30, 50 days with 20, 50, 30 iteration in each window respectively controlling both initial and boundary conditions.

Three 30 years model runs were performed in order to get statistically stable averages. The first one was the reference model run on the grid aligned with the boundary. The second and the third runs were performed on the grid rotated on the   $30^\circ$ with respect to the boundary. Classical free-slip boundary conditions on the staircase shaped boundary have been used in the second run and optimal $\alpha$ have been used  in the third run. 

So far, the major difference in these runs is observed near the western boundary, the zoom of this region only  is shown in the  \rfg{2gyre}. Theoretical middle latitude in this rectangle is equal to the $34^\circ$ latitude (the boundaries are placed at $24^\circ$ and $44^\circ$). The zero SSH line passes at $34.2^\circ$ in the reference and in the optimal experiments but this line starts at $34.8^\circ$ and goes up to the $35.3^\circ$ latitude in the experiment on the rotated grid. That means the point of the separation of the western boundary currents is moved  $0.6^\circ=66$ km to the North and the flux direction is no longer aligned with the latitude. 

The main difference between  \rfg{2gyre}A and  \rfg{2gyre}C consists in the overestimated negative SSH anomaly and in the $1^\circ$ shorter jet. This difference is probably due to insufficient width of the assimilation window. Indeed, we assimilate information about the reference model during 50 days only while  comparing the model runs on 30 years, i.e. more than 200 times longer interval.  However, we make no effort to get more resembling solutions because the purpose of this paper is just to show the potential improvement due to  the boundary conditions optimization. Moreover, a fine justification is not interesting in such an academic configuration as a rectangular box.

\section{Conclusions and perspectives}

We have analyzed the influence of the staircase-like approximation of the boundary  on the model solution and have proposed a way to compensate this influence by optimization of the discretization of the derivatives operators in calculation of the relative vorticity near the boundary. Several experiments carried out with the rectangular box configuration of Nemo model  show the possibility of such a compensation  for a model forced both by a single-gyre and a double gyre surface wind. 

Optimal boundary conditions found in the data assimilation procedure allow us to get the model solution that is rather close to the solution of the reference model. These conditions seem to be independent on the model's resolution, but they depend on all other parameters of the model and have to be determined for each particular configuration. Optimal  model boundary is a curvilinear boundary with the curvature depending on the model configuration and on the approximation of the boundary by the grid. 

Even in the simplest cases,  optimal boundary does not coincide with the presumably known exact one. Considering a real ocean, the coastline  has an extremely  complex, even fractal,  structure and the optimization of its approximation by the data assimilation seems to be a promising way in formulation of an adequate model coastline and boundary conditions. Moreover, in realistic  cases one has to consider a three-dimensional staircase-like approximation of the coast, especially with geopotential vertical coordinate: decreasing ocean depth leads to the formulation of the boundary conditions simultaneously both at the lateral coastline and at the  bottom topography.  Flow interactions with the staircase-like continental slope should also be analyzed in this case.

However, as well as  in \cite{orca-paper}, we can not pretend to solve  a parameter identification problem.  In frames of twin experiments, we assimilate  perfect unperturbed data with no effort to analyze the sensitivity of obtained "optimal" discretizations to possible noise in the observational data. Moreover, 
only a limited number of iterations is allowed in the minimization process leading to the fact that  the convergence is far from even a local minimum of the cost function and, consequently, obtained values are far from the really optimal values for a given configuration. The problem of parameter identification is, of course, a very  interesting, but difficult challenge that must address many mathematical and numerical issues like uniqueness and stability of the identified parameter, convexity of the cost function, etc. All these issues require a particular and detailed study.  The purpose of this paper is to 
 show the result we can potentially get optimizing boundary conditions for realistic models  and which operators and which variables of the model should be optimized in order to compensate model errors due to inappropriate approximation of the boundary.  

The academic configuration used in this paper allowed us to analyze the interaction of the model flow with the lateral boundary and to show that even if we know the position of the exact boundary, we can fail formulating the boundary conditions. The boundary of the rectangle on the $45^\circ$ rotated grid passes exactly by the grid 
nodes, but the model becomes unstable under boundary conditions formulated at these nodes. 

We could think that these instabilities were due to the interpolations ($\overline{\overline{v}^x}^y$ and $\overline{\overline{u}^y}^x$ in \rf{1.1}, \rf{1.2}) in the advection operator and they could be avoided by using some another interpolation scheme. However, it is not evident how to change these interpolations preserving conservation of the mass and the enstrophy. Moreover, we have performed several unsuccessful attempts trying to control the interpolation scheme near the boundary. Indeed, operators  $\overline{u}^y, \overline{v}^x$ are very similar to $\Dy u, \Dx v$, with just different numerical values of coefficients. We can also control the discretization of these interpolations near the boundary by data assimilation. However, similar experiments do not allow to find an optimal scheme: neither the cost function  decreases, nor the boundary current elongates. 

This fact leads us to the conclusion: data assimilation and control of the boundary conditions shows us the easiest, {\bf the most economic way for the model} to bring the solution closer to the observational data. This way may  be different from the way we find the easiest for us. Instead of inclined (with respect to the grid) straight line that was supposed to be the exact  boundary in the experiment with the $45^\circ$ rotated grid, we have got a highly curvilinear boundary that realise the most economic  modification of the boundary conditions from the point of view of the model. 

Obtained "optimal" boundary conditions may violate the physical background of the model conception in cases when there is no other way to bring the model solution closer to observations. Thus, physically doubtful condition of non-zero velocity at the bottom was found by the model to be the easiest way to intensify the surface jet-streams (Gulf Stream and Kuroshio) in the low-resolution global ocean configuration of Nemo known as Orca-2 (see \cite{orca-paper}). Indeed, $2^\circ$ resolution and high lateral viscosity leave no possibility to intensify the jets without external  inflow of energy. Data assimilation just indicates where this inflow must be placed in order to limit it's magnitude. 

On the other hand, issues concerning  stability and conservative properties of the optimal scheme are satisfied almost automatically. The assimilation window must be chosen sufficiently long in order to let potential instabilities be developed within the window. In this case, an unstable solution will not be accepted by data assimilation as an optimal one. Conservative properties, indeed,  may be violated in certain cases, especially when some mass or energy flux is present  in the assimilated data. In this paper, twin data are assimilated ensuring  no  noticeable trend of mass, energy and enstrophy even on 30 years time interval. But, assimilating real data that contain a flux of an integral quantity, we should be ready to add a constraint  in the cost function to ensure the  conservation of an appropriate integral and avoid long-term trends. Thus, we had to add the total  mass conservation requirement in \cite{sw-nl} to compensate the  mass flux in the satellite observations of SSH in the Black sea.

Consequently, if we want to "improve" the model solution, the data assimilation can show us how to do this, but this information should be analyzed from the point of view of agreement with the physical concept of the model. 

The use of the automatic differentiation tool proves to be extremely useful in this study helping us to avoid the huge   coding and debugging work. This fact is appreciated in the situation when we intend to control the distributed parameters rather than the initial condition.

Acknowledgements: 
Author would like to express his gratitude to Julien Le Sommer for helpful discussions on the physical background of the model. 
This work was supported by  the French National Programme LEFE/INSU-GMMC. 
All the contour pictures have been prepared by the Grid Analysis and Display System (GrADS) developed in  the Centre for Ocean-Land-Atmosphere Interactions,   Department of Meteorology, University of Maryland.

%\bibliography{/home/kazan/text/mybibl}

\end{document}